\documentclass[11pt]{article}

[1999/03/29 PSNFSS v.7.2
Times font as default roman
: S Rahtz]

\newcommand \be{\begin{equation}}
\newcommand \ba{\begin{eqnarray}}
\newcommand \ee{\end{equation}}
\newcommand \ea{\end{eqnarray}}

\title{\LARGE Log-periodic route to fractal functions}

\author{S. Gluzman$^{1}$ and D. Sornette$^{1,2.3}$\\
$^1$ Institute of Geophysics and Planetary Physics\\ University of
California Los Angeles, Los Angeles, CA 90095-1567\\
$^2$ Department of Earth and Space Sciences, UCLA\\
$^3$ Laboratoire de Physique de la Mati\`ere Condens\'ee\\ CNRS UMR
6622 and Universit\'e de Nice-Sophia Antipolis, 06108 Nice Cedex 2,
France}

\begin{document}

\maketitle
\begin{abstract}

Log-periodic oscillations have been found to decorate the usual power
law behavior found to describe the approach to a critical point, when
the continuous scale-invariance symmetry is partially broken into 
a discrete-scale invariance (DSI) symmetry. For Ising or Potts spins with 
ferromagnetic interactions
on hierarchical systems, the relative magnitude of the log-periodic corrections
are usually very small, of order $10^{-5}$. In growth processes (DLA), 
rupture, earthquake and financial crashes, they are found of the order of $10\%$.
We offer a ``technical'' explanation for this 4-order-of-magnitude difference based
on the property of the ``regular function'' $g(x)$ embodying the effect of the microscopic
degrees of freedom summed over in a renormalization group (RG) approach $F(x)=
g(x)+\mu^{-1}F(\gamma x)$ of an observable $F$ as a function of a control parameter $x$. 
For systems for which the RG equation has not been derived, the previous equation can
be understood as a Jackson $q$-integral, which is the natural tool for describing
discrete scale invariance. We classify
the `Weierstrass-type'' solutions of the RG into 
two classes characterized by the amplitudes $A_n$ of the power law series expansion. 
These two classes are 
separated by a novel ``critical'' point. Growth processes (DLA), 
rupture, earthquake and financial crashes thus seem to be characterized by oscillatory
or bounded regular microscopic functions that lead to a slow power law decay of $A_n$,
giving strong log-periodic amplitudes. If in addition, the phases 
of $A_n$ are ergodic and mixing, the observable presents
self-affine non-differentiable properties. In contrast, the regular function of
statistical physics models with ``ferromagnetic''-type interactions at
equibrium involves unbound logarithms of polynomials of the control variable that
lead to a fast exponential decay of $A_n$ giving weak log-periodic amplitudes and smoothed
observables. These two classes of behavior might be traced back to the existence or
abscence of ``antiferromagnetic'' or ``dipolar''-type interactions which, when present, make the 
Green functions non-monotonous oscillatory and favor spatial modulated patterns.

\end{abstract}

\pagebreak

\section{Introduction}

The existence of log-periodic oscillatory corrections to the power laws
associated with critical phenomena and, more generally, to observables of systems
endowed with the scale-invariance symmetry has been recognized since the
1960s (see \cite{Sorreview} for a recent review and references therein). The log-periodic
oscillations result from a partial breakdown of the continuous scale-invariance
symmetry into a discrete scale-invariance symmetry, as occurs for instance in 
hierarchical lattices. 

However, for one of the most studied class of models exhibiting these oscillations, 
i.e., Potts model with ferromagnetic interactions 
on hierarchical lattices, the relative magnitude of the log-periodic corrections
are usually very small, of order $10^{-5}$ \cite{DIL}. In contrast,
in growth processes (DLA) \cite{DLA,Growth}, 
rupture \cite{critrup}, earthquakes \cite{SorSam} and financial crashes \cite{SJB,JSL}, 
they are found of amplitude of the order of $10\%$. 

Here, we propose an explanation for this puzzling observation of
an 4-order-of-magnitude difference based on the 
nature of the microscopic interactions of the systems. Within
a renormalization group (RG) approach, an observable at one scale can be related
by a functional relation to the same observable at another scale, with the
addition of the contribution of the degrees of freedom left-over by the
procedure of decimation or of change of scale. This contribution is called
the ``regular part'' of the renormalization group equation of the observable.
For systems for which the RG equation has not been derived, the RG equation can
be understood without reference to the RG
as a Jackson $q$-integral \cite{Jacksonqder}, which is the natural tool 
\cite{Erzan1,ErzanEck} for describing  discrete scale invariance. Here, we do not discuss
the mechanisms by which the continuous scale invariance symmetry is broken
to give discrete scale invariance but rather
present a phenomenological approach based on
the functional RG/Jackson $q$-integral equation.

Using the Mellin transform applied to the formal series solution of the
renormalization group, we identify two broad classes of systems based on
the nature of the decay with order $n$ of the
amplitudes $A_n$ of the power law series expansion of the observable: 
\begin{enumerate}
\item systems with quasi-periodic ``regular part'' and/or with compact support
have coefficients $A_n$ decaying as a power law $A_n \sim n^{-p}$,
leading to strong log-periodic oscillatory amplitudes; if in addition, the phases
of $A_n$ are ergodic and mixing, the observable presents singular properties
everywhere, similar to those of  `Weierstrass-type'' functions.

\item systems with non-periodic ``regular part'' with unbound support have
$A_n$ decaying as an exponential $A_n \sim e^{-\kappa n}$ of their order $n$,
leading to exceedingly small log-periodic oscillatory amplitudes and regular
smooth observables.
\end{enumerate}
We find families of ``regular parts'' which belong to both classes, 
with a ``critical'' transition from the first to the other as a parameter is varied.

A known example of a system of the first class is the $q$-state Potts model 
with {\it antiferromagnetic} interactions \cite{BEE,McKay}. Another example
is the statistics of closed-loop self-avoiding walks per site on a family
of regular fractals with a discrete scale-invariant geometry such as 
the Sieirpinsky gasket \cite{Paul}.
A known example of the
second class is the $q$-state Potts model 
with {\it ferromagnetic} interactions \cite{DIL}. Based on this knowledge and
on the analysis of this paper which underlines the importance of the oscillatory
property of the ``regular part'' or of its compact support to generate strong
log-periodic structure, we conjecture that systems that possess ``antiferromagnetic''-type
or ``dipolar''-type interactions that
introduce modulated spatial patterns in the system
belongs to the first class. This provides a possible explanation for the strong
log-periodic oscillatory amplitudes observed empirically in 
growth processes, rupture, earthquake and financial crashes as we argue in
the concluding section.

The next section 2 introduces the renormalization group with a single control
parameter, its formal solution with the presence of log-periodic corrections
in the presence of discrete scale invariance. Section 3 uses the Mellin transform
to resum the formal series solution of the renormalization group into a power law
series. Section 4 presents the general classification within the
two classes alluded to above in terms
of the leading exponential or power law decay of the coefficients of this power law
expansion. It examines the conditions under which the observable can develop
non-differentiable fractal properties similar to Weierstrass-type functions.
A family of ``regular parts'' is introduced which exhibits a critical transition
between the two classes. Section 5 presents many more examples of both classes.
Section 6 concludes with a qualitative argument explaining why 
rupture and financial systems should belong to the first class. The table 1 offers
a synthesis of the classification in terms of the decay of the coefficients
$A_n$ of the power series expansion of the observable for various choices of the
``regular part'' of the renormalization group.

\section{``Weierstrass-type functions'' from discrete renormalization group equations}

Speaking about a material shape, a mathematical object or function, the
symmetry of scale invariance
refers to their invariance with respect
to changes of scales of observation (see \cite{Dubrulleetal,mybook} for general introductions).
In a nutshell, scale invariance simply means reproducing itself on 
different time or space scales. 
Specifically, an observable $f$ which depends on a ``control'' parameter $x$ is
scale invariant under the arbitrary change $x \to \gamma x$ if
there is a number $\mu (\gamma)$ such that  
\be 
f(x) = {1 \over \mu} f(\gamma x) ~.  \label{oneada}
\ee
Such scale invariance occurs for instance at the critical points $t=t_c$ of systems
exhibiting a continuous phase transition. The renormalization group
theory has been developed to provide an understanding of the emergence of the 
self-similar property (\ref{oneada}) from a systematic scale change
and spin decimation procedure \cite{Goldenfeld}.

Calling $K$ the coupling (e.g. $K=e^{J/T}$ for a spin model where $J$ is interaction 
coefficient and $T$ is the temperature)  and $R$
the renormalization group map between two successive magnification steps, 
the free energy $f$ per lattice site, bond, atom or element obeys the 
self-consistent equation:
\be
f(K)=g(K)+{1\over\mu}f[R(K)]~,    \label{hghaqkq}
\ee
where $g$ is a regular part
which is made of the free energy of the degrees of freedom summed over between
two successive renormalizations, $\mu>1$ is the ratio of the number
of degrees of freedom between two successive renormalizations. In general, this
relationship (\ref{hghaqkq}) is an approximation whose validity requires
the study of the impact of many-body interactions. When these higher-order
interactions can be considered secondary as the scale of description increases
(corresponding to so-called ``irrelevant'' operators),
expression (\ref{hghaqkq}) becomes asymptotically exact at large scales.
For perfectly self-similar problems, for instance for physical systems 
with nearest-neighbor interactions defined
on regular geometrical fractals such as the Cantor set, the
Sierpinsky Gasket, etc., or on regular hierarchical lattices,
expression (\ref{hghaqkq}) is exact at all scales.

It is solved recursively by
\be
f(K)=\sum_{n=0}^\infty {1\over\mu^n}g[R^{(n)}(K)]~,  \label{glrfnglag}
\ee
where $R^{(n)}$ is the $n^{th}$ iterate of the renormalization transformation.
Around fixed points $R(K_c)=K_c$, the renormalization group map can be expanded
up to first order in $K-K_c$ as $R(K) = \gamma (K-K_c)$. Posing $x=K-K_c$,
we have $R^{(n)}(x)=\gamma^n x$ and
the solution (\ref{glrfnglag}) becomes
\be
f(x)=\sum_{n=0}^\infty {1\over\mu^n}g[\gamma^{n} x]~.  \label{glrfnaaglag}
\ee
In principle, (\ref{glrfnaaglag}) is only applicable sufficiently ``close''
to the critical point $x=0$, that the higher-order terms in the expansion 
$R(K) = \gamma (K-K_c)$ can be neglected. The effect of nonlinear corrections terms for 
$R(K)$ have been considered in \cite{BEE,DIL}.

The form (\ref{glrfnglag}) or (\ref{glrfnaaglag}) has not been derived rigorously
for growth, rupture and other out-of-equilibrium processes alluded to above, even
if there are various attempts to develop approximate RG descriptions on specific
models of these processes.
It may thus seem a little premature to use this discrete renormalization group 
description for these systems. Actually, expression (\ref{glrfnaaglag}) can
be obtained without any reference to a renormalization group approach: as soon
as the system exhibits a discrete scale invariance, the natural tool is provided
by $q$-derivatives  \cite{ErzanEck} from which it is seen that expression
(\ref{glrfnaaglag}) is nothing but a 
Jackson $q$-integral \cite{Jacksonqder}
of the function $g(x)$, which constitutes the natural generalization
of regular integrals for discretely self-similar systems \cite{ErzanEck}.
The way the Jackson $q$-integral is related to the free energy of a spin
system on a hierarchical lattive was explained in \cite{Erzanderivation}.

In the mathematical literature, the function (\ref{glrfnaaglag}) is called
a {\it Weierstrass-type function}, to refer to the introduction by K. Weierstrass
of the function \cite{Weierstrass}
\be
f_W=\sum_{n=0}^\infty b^n \cos[a^{n} \pi x]~,  \label{wiuelag}
\ee
corresponding to the special case
$\mu=1/b$, $\gamma=a$ and $g(x) = \cos[\pi x]$. To the surprise
of mathematicians of the 19th century, Weierstrass
showed that the function (\ref{wiuelag}) is continuous 
but differentiable nowhere, provided
$0 < b < 1, a>1$ and $ab>1+{2 \over 3}\pi$. Note that, in the 
context of the renormalization group of critical phenomena, the condition $a=\gamma>1$
implies that the fixed point $K_c$ is unstable. 
Hardy was able to improve later on the last bound 
and obtain that the Weierstrass function (\ref{wiuelag}) is non-differentiable
everywhere as soon as $ab >1$ \cite{Hardy}. In addition, Hardy showed that it
satisfies the following Lipschitz condition (corresponding to self-affine 
scaling) for $ab>1$,
which is much more than just the statement of non-differentiability:
\be
f_W(x+h)-f_W(x) \sim |h|^m~, ~~~{\rm for~all}~x~~{\rm where}~~  m={\ln [1/b]/\ln a}~.
\label{Lipschitz}
\ee
Note that for $ab>1$, $m<1$, expression (\ref{Lipschitz}) shows that $f_W(x+h)-f_W(x) \gg |h|$ 
for $h \to 0$.
As a consequence, the ratio $[f_W(x+h)-f_W(x)]/h$ has no limit which recovers
the property of non-differentiability. Continuity is obvious from the fact that 
$f_W(x+h)-f_W(x) \to 0$ as $h \to 0$ since $m>0$. For the border case $a=b$ 
discovered by Cellerier before 1850, $f_W$ is not non-differentiable in a strict sense
since it possesses infinite differential coefficients at an everywhere dense set of 
points \cite{Singh}. Richardson is credited with the first 
mention of the potential usefulness for 
the description of nature of the continuous everwhere non-differentiable
Weierstrass function \cite{Richardson}. Shlesinger and co-workers
\cite{Shlesinger} have previously noticed and studied the correspondence between 
(\ref{glrfnaaglag}) and Weierstrass function.

If one is interested in the non-regular (or non-analytic)
behavior only close to the critical point $x=0$, the regular part can be dropped and
the analysis of (\ref{oneada})
is sufficient. It is then easy to show that the most general solution of (\ref{oneada}) is
(see \cite{Sorreview} and references therein)
\be
f(x) = x^m ~P\left({\ln x \over \ln \gamma}\right)~,
\label{gensolcrit}
\ee
where
\be
m={\ln \mu \over \ln \gamma}~,   \label{mvalue}
\ee
and $P(y)$ is an arbitrary periodic function of its argument $y$ of period $1$. Its specification 
is actually determined by the regular part $g(x)$ of the renormalization group equation,
as shown for instance in the explicit solution (\ref{glrfnaaglag}). The scaling law
$f(x) \sim x^m$ implied by
(\ref{gensolcrit}) is a special case of (\ref{Lipschitz}) obtained by putting $x=0$ and
replacing $h$ by $x$ in (\ref{Lipschitz}).

The Laplace transform $f_L(\beta)$ of $f(x)$ defined by (\ref{glrfnaaglag}) also obeys a 
renormalization equation of the type (\ref{hghaqkq}). Denoting $g_L(\beta)$ the Laplace
transform of the regular part $g(x)$, we have
\be
f_L(\beta) = \sum_{n=0}^\infty {1\over (\mu \gamma)^n} g_L[\beta/\gamma^{n}]~,  
\label{gaalrfnaaglag}
\ee
and
\be
f_L(\beta) = g_L(\beta) + {1 \over \mu \gamma} f_L \left({\beta \over \gamma} \right)~.
\label{gnhgqllqa}
\ee
The general solution of (\ref{gnhgqllqa}) takes the same form as (\ref{gensolcrit}):
\be
f_L(\beta) = {1 \over \beta^{1+m}}~P_L\left({\ln \beta \over \ln \gamma}\right)~,
\ee
where $P_L(y)$ is an arbitrary periodic function of its argument $y$ of period $1$.

\section{Reconstruction of ``Weierstrass-type functions'' from power series expansions}

Following \cite{DIL,salsor}, we use the Mellin transform to obtain a power law
series representation of the {\it Weierstrass-type function} (\ref{glrfnaaglag}).
The Mellin transform is defined as
\be
\hat{f}(s)\equiv \int_0^\infty x^{s-1}f(x)dx~.  \label{Mellintransdef}
\ee
The Mellin transform (\ref{Mellintransdef}) provides a reconstruction of the 
infinite sum of the {\it Weierstrass-type function} (\ref{glrfnaaglag})
as a sum of power law contributions, $A_n x^{-s_n}$, with ``universal'' 
complex exponents $s_n$ determined only by properties of the hierarchical construction and
not by the function $g(x)$, with amplitudes $A_n$
controlled by the form of the regular part $g(x)$. 
These ``non-universal'' amplitudes in turn control the shape of the function 
$f(x)$, its differentiability or non-differentiability as well as its self-affine
(fractal) properties, as we shall describe in the sequel.

The Mellin transform of (\ref{glrfnaaglag}) reads
\be
\widehat{f}(s)=\frac{\mu \gamma ^s}{\mu \gamma ^s-1}\ \widehat{g}(s)~, \label{mellintrfund}
\ee
where $\widehat{g}(s)$ is the Mellin transform of $g(x)$.
The inverse Mellin transformation of $\widehat{f}(s)$,
\be
f(x)=\frac 1{2\pi i}\int_{c-i\infty }^{c+i\infty }\widehat{f}(s)x^{-s}ds~, 
\ee
allows us to reconstruct $f(x)$ as a new expansion in singular as well as
regular powers of $x$ in order to unravel its self-similar properties. 
Indeed, the usefulness of the Mellin transform is that power law behaviors
spring out immediately from the poles of $\widehat{f}(s)$, using Cauchy's theorem.

In inverting the Mellin transform, we have two types of poles. The poles
of the Mellin transform $\hat{g}$ of the analytical function $g(x)$
occur in general at integer values 
and contribute only to the regular part $f_r(x)$
of $f(x)$, as expected since $g(x)$ is a regular
contribution. The poles of the first term $\frac{\mu \gamma ^s}{\mu \gamma ^s-1}$
in the r.h.s. of (\ref{mellintrfund})
stem from the infinite sum over successive embeddings of scales and occur at
$s=s_n$ where
\be
s_n= -m +i \frac{2\pi}{\ln \gamma }\ n~,    \label{expgenfrac}
\ee
and $m$ is given by (\ref{mvalue}).
Their amplitude $A_n$ is obtained by applying Cauchy's theorem
and is given by the residues
\be
\lim_{s\rightarrow s_n}\frac{s-s_n}{\mu \gamma ^s-1}~\widehat{g}(s)=
\frac{\exp (-2\pi n\ i) }{\ln \gamma }~\widehat{g}(s)=\frac{\widehat{g}(s)}{\ln \gamma}~.
\ee
The resulting expression for $f(x)$ is
\be
f(x)=f_s(x)+f_r(x)~,   \label{gensolsumsinreg}
\ee
where the singular part $f_s(x)$ is given by
\be
f_s(x)=\sum_{n=0}^\infty A_n\ x^{-s_n}~,   \label{sinsumgen}
\ee
and
\be
A_n=\frac{\widehat{g}(s_n)}{\ln \gamma}~.   \label{amplgensing}
\ee
This approach is similar to the one developed in \cite{lapidus} for ``fractal strings''
$\eta$ (for instance, the complementary of the triadic Cantor set
is a special fractal string). 
Their fractal properties are fully characterized by the introduction
of the ``geometric zeta function'' $\zeta_{\eta}(s)$, which
can be shown to be nothing but the Mellin transform of the measure defined on
the fractal string (see \cite{lapidus} page 73). 
In particular, the poles of $\zeta_{\eta}(s)$
give the complex fractal dimensions of the fractal strings, similarly to the
role played here by the complex exponents $s_n$ defined by (\ref{expgenfrac}).

The regular part $f_r(x)$ of $f(x)$ defined in (\ref{gensolsumsinreg}) is
generated by the poles of $\widehat{g}(s)$ if any, located at $s=-n$, $n=0,1...$. 
The residues of these poles give the coefficients $B_n$ of the expansion of
the regular part as follows:
\be
f_r(x)=\sum_{n=0}^\infty B_n\ x^n~.   \label{reuughnla}
\ee

\section{Classification of ``Weierstrass-type functions''}

\subsection{Classification \label{classecffr}}

The representation (\ref{sinsumgen}) offers a classification of 
``Weierstrass-type functions'' as follows.
We will work in the class of $g(x)$ (not covering of course all possible types
of behavior of $A(n)$) where
the coefficients $A_n$ can be expressed as the product of an exponential decay
by a power prefactor and a phase
\be
A_n  = {1 \over \ln \gamma}
{1 \over n^p}~e^{- \kappa n}~e^{i \psi_n},~~~~{\rm for~large}~~n~,  \label{classangener}
\ee
where $p$, $\kappa  \geq 0$ and $\psi_n$ are determined by the form of $g(x)$ and the values of 
$\mu$ and $\gamma$. This class is broad enough to include many physically interesting shapes of
$g(x)$ as will be illustrated at length below.

\subsubsection{Justification of the classification}

The parameterization (\ref{classangener}) can be seen to result from 
very general theorems on the Mellin transform \cite{Sidorov,Oberhett}.
Let us assume that the function $g(x)$ defined for $x>0$ is continuous
and satisfies the following conditions
\be
\left| g(x\right| \leq c_1 x^\alpha ~, ~~~~~~0<x\leq 1;\qquad \left|
g(x\right| \leq c_2\ x^\beta ~,~~~~~~1\leq x<\infty ~,  \label{hghngklda}
\ee
where $\alpha >\beta$. Then, its Mellin transform is a regular (differentiable) function
inside the strip $-\alpha <Re(s)<-\beta$, One should also bear in mind
that $Re(s)<0$ because of the constraints imposed by the very formulation of
the problem. All functions that we shall consider below as examples belong to
the class of continuous functions satisfying slightly more restricted conditions 
\cite{Sidorov} such as (\ref{hghngklda}) with $\alpha >0$ and $\beta =0$.
As a consequence, their Mellin transform is regular for $-\alpha <Re(s)<0$. 
For instance, $g(x)=cos(x)-1$ corresponds to $\alpha =1$ and $\beta=0$. The same conditions
apply to $log(1+x)$ and $exp(-x)-1$. For the stretched exponential function $exp(-x^h)-1$
with $h>0)$, we have $\alpha=h$ and again $\beta =0$.

We are interested, in $\widehat{g(s_n)} = \widehat{g}(-m+in\omega)$,
particularly as $n$ goes to infinity. The general condition which is usually
imposed on this quantity in order to ensure the existence of its inverse Mellin
transform is \cite{Oberhett}
\be
\widehat{g}(-m+in\omega )\rightarrow 0,~~~~{\rm as}~~~n \rightarrow +\infty ~.
\ee
Again, $A_n$ must be designed in such a way that it satisfies this condition
automatically.

Let us consider some simple but vivid examples, intended to illustrate how
a power-law and exponential decay of $A_n$ as a function of $n$
emerges from simple functions
satisfying the conditions stated above. We also note that, when $g(t)$ possesses
discontinuities of the first kind, it still yields the dependence (\ref{classangener})
of $A_n$ as a function of $n$.
Maybe the simplest function leading to $A_n$ with a power-law decay is
\be
g(x)=0,~~~~~ 0<x<1 ~~~{\rm and}~~~~g(x)=-1~,~~~~~ 1<x<\infty~,
\ee
which leads to $\widehat{g(s)}=\frac{1}{s}$, which is regular within the
strip $\infty <Re(s)<0$. The corresponding $A_n$ decays in amplitude as
$n^{-1}$ for large $n$. Different strip geometries lead to the same power-law
decay of $A_n$, for instance
\be
g(x)=x^a,~~~~~ 0<x<1~~~~{\rm and}~~~~g(x)=0,~~~~ 1<x<\infty~,
\ee
with Mellin transform $\widehat{g(s)}=(s+a)^{-1}$ with $-a<Re(s)<0$.
Let us also consider
\be
g(x)=0~,~~~~~ 0<x<1~~~~{\rm and}~~~~g(x)=-x^a~,~~~~ 1<x<\infty~,
\ee
which leads to a similar Mellin transform $\widehat{g(s)}=(s+a)^{-1}$
but a different strip
geometry $Re(s)<-a$. The slightly more
complicated example of a continuous function composed of power laws
\be
g(x)=(b-a)^{-1}x^a~,~~~~~ 0<x<1~~~~~{\rm and}~~~~~g(x)=(b-a)^{-1}x^b~,~~~~ 1<x<b~,
\ee
leads to $\widehat{g(s)}=(s+a)^{-1}(s+b)^{-1}$ with $-a<Re(s)<-b$ and the 
amplitude of $A_n$
decaying as $n^{-2}$. The analysis of these examples and of their Mellin
transforms at $s=s_n$ demonstrate that particulars of the strip geometry in the
variable $s$
are not important when one is concerned with the large $n$ asymptotic behavior
of $\widehat{g}(s_n)$. The asymptotic power decay of $A_n$ as a function of $n$
can be dominated by an exponential decay, as we shall see in more details below.
For instance, the continuous function formed by compounded power-laws
$g(x)=\frac 1\pi \ x(1+x)^{-1}$ leads to $\widehat{g(s)}=-\csc (\pi
s),\qquad -1<Re(s)<0$, yielding $A_n$ decaying as $exp(-n)$ as
$n \to +\infty$.

Violations of the parameterization (\ref{classangener}) regarding $A_n$
occur when the conditions of the
theorem \cite{Sidorov} are changed, e.g. when the argument $x$ is replaced by,
say, $\ln(1/x)$ or when singularities are introduced to the function $g(x)$.
This can be seen from equation 
(\ref{amplgensing}) which shows that $A_n$ is proportional to the Mellin transform of
$g(t)$ expressed at $s=s_m=-m+in\omega$ where 
\be
\omega =\frac{2\pi }{\ln \gamma }~.   \label{guroww}
\ee
Posing $u=\ln x$, the Mellin transform becomes a Fourier transform
\be
A_n = {1 \over \ln \gamma}~ \int_{-\infty}^{+\infty} du ~ G(u) ~ e^{i \omega n u} ~,   
\label{gjnbflwa}
\ee
where
\be
G(u) \equiv e^{u(1-m)} g(e^u)~.
\ee
It is clear that, by a suitable choise of $g(x)$, any dependence of $A_n$ can be obtained.
For instance, for 
\be
G(u) = u^{-3/2} ~e^{-a/u}~,  \label{giwgjgwas}
\ee
we obtain $A_n \sim e^{-\sqrt{2an}}~\cos[\sqrt{2an}]$, which exhibits an 
oscillatory stretched-exponential decay intermediate between
the exponential ($\kappa>0$) and pure power law decay ($\kappa=0$) of (\ref{classangener}).
However, the choice (\ref{giwgjgwas}) corresponds to a rather special choice for 
\be
g(x) = {(\ln x)^{-3/2}~e^{-a/\ln x} \over t^{1-m}}~.
\ee
In this case, $g(x) \to +\infty$ for $x \to 0$, and this case is outside the
domain of validity (\ref{hghngklda}) of the theorem \cite{Sidorov,Oberhett}.
Consider also the following example
\be
g(x)=\pi^{-1/2}\exp \left( -\frac{\ln (1/x)^2}4\right) ~,
\ee
leading to $\widehat{g(s)}=\exp (s^2)$,
valid for arbitrary $s$, which leads to $A_{n}$ with amplitude
decaying as $exp(-n^2)$ as $n \to \infty$. 
This example is also characterized by a pathological behavior 
for $x \to +\infty$ of $g(x)$ which diverges faster than any power law. 
Another pathological example is
\be
g(x)=1/2\ \pi ^{-1/2}\ \cos \left( (1/4) \ln(1/x)^2-\pi/4 \ \right)~,
\ee
leading to $\widehat{g(s)}=\cos  s^2$, 
valid for arbitrary $s$, which yields $A_{n}$ with an amplitude growing
as $\exp(n)$ as $n \to +\infty$. This violates the
condition on the Mellin transform given in \cite{Oberhett}.

The existence of discontinuities of $g(x)$, as 
one might expect from the theorem \cite{Sidorov,Oberhett},
also violates the parameterization (\ref{classangener}) of $A_n$. 
Consider $g(x)=-\frac 1\pi \ x^{1/2}(1-x)^{-1}$ with an integrable singularity, 
which gives $\widehat{g(s)}=\tan (\pi s),\qquad -1/2<Re(s)<0$ and
$A_n$ with an amplitude
bounded from below by a constant as $n \to +\infty$. This absence of decay
allows us to reject this type of function, since a decay of $A_n$ is
required by \cite{Oberhett},
In constrast, a logarithmic singularity, as for instance in $g(x)=\frac
1\pi \ \log \left| \frac{1+x}{1-t}\right|$, is allowed. In this case,
this gives $\widehat{g(s)} =s^{-1}\tanh (\pi s),\ -1<Re(s)<0$ and
the amplitude of $\widehat{g(s_n)}$ exhibits
periodic modulations as $n \to +\infty$ as
$n^{-1}\left( \sinh {}^2(\frac 12\pi m)+\cos {}^2(\frac 12\pi
\omega n)\right) ^{-1}\left( \sin (\pi \omega n)+i\sinh (\pi m)\right)$.
Another example with the logarithmic function (\ref{bgpgfqqf}) discussed
below gives a power law decay with a logarithmic correction as shown by (\ref{gnhnllsls})
due to the presence of the singularity.

In conclusion, as long as the
conditions of theorem \cite{Sidorov} on Mellin transforms hold, the 
dependence of $A_n$ as $n \to +\infty$ given by (\ref{classangener})
will hold as well. Violations of the theorem due to a change of variable or
to the presence of simple poles lead either to a faster decay or to a non-decaying
$A_n$. Allowing for logarithmic singularities within $g(x)$ brings in
logarithmic or oscillatory corrections to $A_n$ as a function of $n$.

\subsubsection{Beyond the linear approxiation of the renormalization group map}

The asymptotic expansion (\ref{classangener}) uses the linear approximation
$R^{(n)}(x)=\gamma^n x$ that allows us to transform the general solution
(\ref{glrfnglag}) into the ``Weierstrass-type function'' (\ref{glrfnaaglag}).
As we said, (\ref{glrfnaaglag}) is only applicable sufficiently ``close''
to the critical point $x=0$, such that the higher-order terms in the expansion 
$R(x) = \gamma x$ can be neglected. The linear
approximation of $R^{(n)}(x)=\gamma^n x$ is bound however to become incorrect
as $n$ becomes very large, i.e., in the region determining the singular behavior.
As discussed in \cite{DIL,salsor}, the crucial property missed by the linear approximation
is that $f(x)$ is analytic only in a sector $|{\rm arg}~ x|<\theta$ while we treated it
as analytic in the cut plane $|{\rm arg}~ x|<\pi$. This 
implies that the exponential contribution $e^{- \kappa n}$ of the true asymptotic decay
of the amplitudes of successive log-periodic harmonics is 
slower than found from the linear approximation, and goes as $e^{- \kappa \theta n}$. The angle
$\theta$ depends specifically on the flow map $R(x)$ of the discrete renormalization
group \cite{DIL} and is generally of order $1$. Our classification in two sets
$\kappa=0$ and $\kappa\neq 0$ is not modified by this subtlety.
Here, we shall consider only 
the ``Weierstrass-type functions'' (\ref{glrfnaaglag}) and will revisit the 
impact of nonlinear terms of the renormalization group map in a future communication.

\subsubsection{$\kappa >0$: $C^{\infty}$-differentiability}

The general solution (\ref{gensolcrit}) remains true for any choice of the regular part
$g(x)$ with the exponent $m$ given by (\ref{mvalue}). This implies that there will 
always be an order of differentiation sufficiently large such that its becomes
infinite at $x=0$ \cite{BEE}. This is the crux of the argument on the existence of the 
singularity at $x=0$. Here, we investigate the differentiability of
$f(x)$ for non-zero values of $x$, i.e., away from the unstable critical point $x \to 0$.
Expression (\ref{sinsumgen}) with (\ref{expgenfrac}) 
provides a direct way for understanding the origin of the singular behavior at $x \to 0$,
as $x^m$ is in factor of an infinite sum of oscillatory terms with log-periodic oscillations 
condensing geometrically as $x \to 0$.

When $\kappa  >0$, the modulus of $A_n$ decay exponentially fast to zero. Hence,
$f(x)$ is differentiable at all orders. This can be seen from the fact that
\be
\frac{d^{\ell}f_s(x)}{dx^{\ell}}=\sum_{n=0}^\infty 
(-s_n)(-s_n-1)...(-s_n-\ell+1) \ A_n\ x^{-s_n-\ell}   \label{dervsumseries}
\ee
is absolutely convergent for any order $\ell$ of differentiation. 
Taking into account that $|x^{-s_n-\ell}|=x^{m-\ell}$ is independent of $n$
and can be factorized,
the $n$th term in the sum has an amplitude bounded
by a constant times $n^{\ell} \exp[-qn]$ since $(-s_n)(-s_n-1)...(-s_n-\ell+1)$
is bounded from above by a constant times $n^{\ell}$. The sum is thus controlled
by the exponentially fast decaying coefficients $A_n$ and converges 
to well-defined values for any $\ell$. As a consequence of the exponential decay of the
coefficients $A_n$, the log-periodic oscillations are extremely small.

Another obvious way to ensure differentiability even when $\kappa =0$
(see next section) is to truncate the number $n$ of powers $x^{-s_n}$
in the sum (\ref{sinsumgen}) to a finite value:
\be
f_s^{(N)}(x)=\sum_{n=0}^N A_n\ x^{-s_n}~,   \label{sinsffdumgen}
\ee
An example with $N=1, 2, 3$ is shown in figure 1 for the
Weierstrass function ($\alpha =\pi /2$ and $p=m+{1 \over 2}$). For
$N=1$, the real part $f_s^{(1)}(x)$ is given by
\be
f_s^{(1)}(x) = a_0\left[ 1+\frac{A_{n=0}}{a_0}x^m+\frac{\left| A_{n=1}\right| }{a_0}%
x^m\cos (\omega \ln (x)+\varphi )\right] ~,~~~~~~ a_0=\frac \mu {\mu -1}~, 
\label{logperiodiccase}
\ee
where $\omega$ is given by (\ref{guroww})
and
\be
\varphi =\arctan \left( \frac{Im\left( A_{n=1}\right) }{Re\left( A_{n=1}\right) }%
\right) +k\pi ,\quad k=0,\pm 1...
\ee
This expression (\ref{logperiodiccase}) is based on the singular part (\ref{sinsumgen}) 
of the Mellin decomposition of the DSI equation (\ref{glrfnaaglag}). It applies
not only to the Weierstrass function but to any function of the form (\ref{glrfnaaglag}).
Keeping only the first two terms
recovers exactly the log-periodic formula introduced in the 
study of precursors of material failure \cite{Anifrani,canonical,critrup}, 
of earthquakes precursors
\cite{SorSam,Salsamsor,Johsalsor} and of precursors of financial crashes \cite{SJB,JSNasdaq}.

\subsubsection{Critical behavior and non-differentiability}

Expression (\ref{sinsumgen}) with (\ref{expgenfrac}) and (\ref{classangener}) shows
that $f_s(x)$ has the same differentiability properties as
\be
\sum_{n=1}^{+\infty} {1 \over n^p}~e^{i \psi_n}~x^{m-i2\pi n/\ln \gamma}~.
\ee
Changing variable $x \to y=\ln x/\ln \gamma$, this reads
\be
e^{y \ln \mu} \sum_{n=1}^{+\infty} {1 \over n^p}~e^{i(-2\pi n y +\psi_n)}~.
\ee
With respect to the differentiability property, it is sufficient to study the
real part of the infinite sum which reads
\be
K_{p,\{\psi_n\}}(y) =\sum_{n=1}^{+\infty} {\cos[2\pi n y +\psi_n] \over n^p}~.  \label{jgnagn}
\ee

This expression allows us to recover some important results in the case where the
phases $\psi_n$ are sufficiently random so that the numerators $\cos[2\pi n y +\psi_n]$
take random uncorrelated signs with zero mean. Then, the sum $K_{p,\{\psi_n\}}(y)$ 
truncated at $n=T$ has the same convergence properties for $T \to \infty$ as 
\be
X(T) = \int_1^T {dW_t \over t^p}~,    \label{hgjwq}
\ee
where $dW_t$ is the increment of the continuous white noise Brownian motion of zero mean and 
correlation function $\langle dW_t dW_{t'} \rangle  = \delta(t-t) dt$ where
$\delta$ is the Dirac function. We get $\langle X(T) \rangle =0$ and its
variance is
\be 
\langle [X(T)]^2 \rangle = \int_1^T  \int_1^T \langle dW_t dW_{t'} \rangle  t^{-p} t'^{-p} 
= \int_1^T  {dt \over t^{2p}}~,
\ee 
which is finite for $T \to +\infty$ if $p>1/2$. This entails
the convergence for $p>1/2$ of the infinite series (\ref{jgnagn}) for most
phases $\psi_n$ which are sufficiently ergodic and mixing.
We thus expect that $K_{p,\{\psi_n\}}(y)$
and as a consequence $f_s(x)$ are continuous functions for $p>1/2$. We can proceed
similarly for studying their $\ell$'s derivative. 
With respect to the convergence property, taking the $\ell$'s derivative has the effect
of changing $p$ into $p-\ell$ in (\ref{hgjwq}). We thus expect $K_{p,\{\psi_n\}}(y)$
and as a consequence $f_s(x)$ to be differentiable of order $\ell$ for $p>\ell+1/2$.

We conjecture the following.

\vskip 0.5cm
{\bf Conjecture on conditions for nondifferentiability from the singular power law
expansion of ``Weierstrass-type functions''}:
provided that (1) $\kappa =0$ and (2) the phases $\psi_n$ are ergodic and, using
a generalization of Hardy's condition $ab>1$ for the Weierstrass
function, the smallest order $\ell_{\rm min}$ of differentiation 
of $f_s(x)$ (defined by (\ref{sinsumgen}) with exponents $s_n$ given by
(\ref{expgenfrac}) with (\ref{mvalue})) which does not exist is such that 
\be
{1 \over 2} < p - \ell_{\rm min} < {3 \over 2}~,
\ee
i.e., 
\be
\ell_{\rm min} = {\rm Int}[p-{3 \over 2}]
\ee
is the integer part of $p-{3 \over 2}$. In particular, with ergodic phases $\psi_n$ of
zero mean, the function $f_s(x)$ is nondifferentiable for $p<3/2$.

\vskip 0.5cm
It follows from Lebesgue's theorem on continuous functions of bounded variations that
a non-differentiable function is not a function of bounded variation. Therefore,
a non-differentiable function is everywhere oscillating and the length of arc between
any two points on the curve is infinite \cite{Singh}. This explains the observation
below that the regular part $g(x)$ must contain oscillations or must exhibit a 
compact support (so that it has a discrete Fourier series) in order for $f(x)$ 
to be non-differentiable or for some of its derivatives to be non-differentiable.
Actually, Weierstrass-type functions (\ref{glrfnaaglag}) are believed to have
the same Hausdorff dimension $2-m$ as the Weierstrass function (\ref{wiuelag})
for arbitrary regular part $g(x)$, as long as it is a bounded 
almost periodic Lipschitz function of order $\beta>m$ \cite{HuLau}. The examples
organized below in two classes illustrate and make precise this condition on $g(x)$. We indeed
find that non-differentiability occurs at a finite order of differentiation
only for functions $g(x)$ which are periodic or with compact support. 

It appears however that there is not yet a general understanding whether there
exists a necessary and sufficient condition for the differentiability of a function
on an interval. It is well-known that continuity is necessary for differentiability
but is not sufficient as shown by the Weierstrass function and other examples above.
The restriction of bounded variations has also proved insufficient: although
a continuous function must possess a differential coefficient almost everywhere, yet
there are examples of such functions which do not possess differential
coefficients at unenumerable everywhere dense sets of points \cite{Singh}.

\subsection{General condition for $\kappa=0$}

Let us consider a regular function $g(x)$ which is either periodic with 
period $X$ or with 
compact support over the interval $[0,X]$ and
zero outside. It can then be expanded as a Fourier series
\be
g(x) = {a_0 \over 2} + \sum_{k=1}^{+\infty} \left[a_k \cos (2\pi kx/X) + 
b_k \sin (2\pi kx/X) \right]~,   \label{bhggna}
\ee
where $a_0, a_1, b_1, ..., a_k, b_k,...$ are arbitrary real numbers.

The behavior of the coefficients $A_n$ is controlled by the Mellin
transform $\widehat{g}(s_n)$ of $g(x)$ as shown by (\ref{amplgensing}). 
For $g(x)$ periodic with zero mean, $a_0=0$ and
\be 
\widehat{g}(s_n) = \left({X \over 2\pi}\right)^{s_n}
\left[\widehat{\cos}(s_n) \sum_{k=1}^{+\infty} {a_k \over k^{s_n}} 
+ \widehat{\sin}(s_n) \sum_{k=1}^{+\infty} {b_k  \over k^{s_n}}\right]~,
\label{hgnvallq}
\ee
where $\widehat{\cos}(s)$ and $\widehat{\sin}(s)$ are the Mellin
transform of $\cos x$ and $\sin x$. 
Now, a general theorem on the Fourier series of periodic functions tells us
that, if $g(x)$ has continuous derivatives up to order $r-1$ included and
if the derivative of order $k$ obeys the Dirichlet conditions, then the
coefficients $a_k$ and $b_k$ decay for large $k$ as $1/k^{r+1}$, i.e., there
is finite $M'>M>0$ such that
$M'/k^{r+1} >|a_k| > M/k^{r+1}$ and $M'/k^{r+1} >|b_k| > M/k^{r+1}$. 
If $g(x)$ is discontinuous
at a discrete set of points, this corresponds to taking $r=0$ in the previous
formula. The Dirichlet conditions
are: (i) $g(x)$ is continuous or possess only a finite number of 
discontinuities; (ii) each point of discontinuity $x_d$ is a 
discontinuity of the first kind, i.e., it is such that the limits
to the left
$g(x \to x_d^-)$ and to the right $g(x \to x_d^+)$ are finite; (iii)
the interval $[0,X]$ can be divided into a finite set of subintervals
on each of which $g(x)$ is monotonous. 

We can thus write
\be 
2M \left({X \over 2\pi}\right)^{s_n}
\widehat{\sin}(s_n) \sum_{k=1}^{+\infty} {k^{-i n \omega} \over k^{r+1-m}} <
\widehat{g}(s_n) < 2M' \left({X \over 2\pi}\right)^{s_n}
\widehat{\sin}(s_n) \sum_{k=1}^{+\infty} {k^{-i n \omega} \over k^{r+1-m}} ~.
\ee

The sum
\be
 \sum_{k=1}^{+\infty }{ k^{-(r+1-m+in\omega )}}~,
\label{nhgnkfvf}
\ee
is nothing else but the celebrated zeta-function $\zeta (y)$ of
Riemann \cite{Titchmarsh,Edwardszeta}, with the correspondence $y=\sigma +it,\ 
\sigma =r+1-m,\ t=n\omega$. It is
known \cite{Titchmarsh,Edwardszeta} that 
$|\zeta (\sigma +it)| \leq C_{\sigma} ~(|t|+1)^{1/2-\sigma}$ for $\sigma <0$,
where $C_{\sigma}$ decreases like $(2 \pi e)^{\sigma -1/2}$ for $\sigma \to -\infty$,
and it does not satisfy a better estimate in this half-plane. 
For  $0 \leq \sigma \leq 1$ (corresponding to  $0 \leq m \leq 1$ and $p=0$,
$\zeta (\sigma +it)| \leq K t^{(1-\sigma)/2}\ln(t))$ uniformly for some constant $K$.
However, we need the behavior of $\zeta (\sigma +it)$ for $\sigma = r+1-m>0$. It is
obtained by using
the relation $\zeta(s) = 2^s \pi^{s-1} \sin{\pi s \over 2} \Gamma(1-s) \zeta(1-s)$, which
which can be separated into two parts that can be evaluated. Namely, $\sin
(\frac \pi 2s)\ \Gamma (1-s)$, re-casted in the variable $z=1-s$, takes the familiar
form,$\cos(\frac \pi 2z)\ \Gamma (z),$ which behaves for large $n$ as 
$n^{-r+m-1/2}$. The other term $\zeta (1-s)=\zeta (z)$ can be evaluated using the
expression presented above for the zeta-function of an argument with negative real part
(in our case for negative $1-\sigma$ and large $n$),
$\zeta (z) \leq C\ \left( \left| n\right| +1\right) ^{1/2-(1-\sigma)}=C(\left|
n\right| +1)^{1/2+r-m}.$ Therefore, the product of these two
terms is of the order of C and the whole sum decay is slower than
exponential.

This shows that the sum (\ref{nhgnkfvf}) is of order $O(1/n^{r+(1/2)-m})$
and thus $\widehat{g}(s_n)$ is asymptotically a negative power of $n$ for large $n$.
This demonstrates that any periodic continuous function $g(x)$ leads to a power law decay
for $A_n$ as a function of $n$. The continuity of $g(x)$ implies that $r \geq 1$, 
ensuring that $r+(1/2)-m>0$ for $m<1$.

The same approach can be used for $g(x)$ not periodic but defined on 
a compact support $[0,X]$. The discrete Fourier series expansion (\ref{bhggna})
still holds for $x \in [0,X]$ while $g(x)=0$ for $x$ outside. A similar
expression to (\ref{hgnvallq}) then holds in which $a_0 \neq 0$ in 
general and in which the Mellin transforms 
$\widehat{\cos}(s)$ and $\widehat{\sin}(s)$ are defined over the 
interval $[0,X]$.

\subsection{Bifurcation from wild to smooth ``Weierstrass-type functions'': an 
example using damped oscillators for the regular part of the renormalization group 
equation}

As a first example, let us consider the regular part $g(x)$ of the renormalization group 
equation defined as
\be
g(x)= e^{-\cos (\alpha )~x}~~ \cos \left( x\ \sin (\alpha)\right)~,~~~
~~{\rm with}~~\alpha \in \left[ 0,\frac \pi 2\right]~,  
\label{exbifreg}
\ee 
The parameter $\alpha$ quantifies the relative strength of the oscillatory
structure of $g(x)$ versus its ``damping'': for $\alpha=\pi/2$, 
(\ref{glrfnaaglag}) with (\ref{exbifreg}) recovers the
initial function (\ref{wiuelag}) introduced by Weierstrass with
$b=1/\mu$, $a=\gamma$ and $\cos(\pi x)$ replaced by $\cos(x)$;
for $\alpha=0$, $g(x)=\exp[-x]$ has no oscillation anymore and corresponds
to a pure exponential relaxation considered in \cite{Klafteretal}.

Plugging (\ref{exbifreg}) in (\ref{glrfnaaglag}) gives
\be
f(x)=\sum_{n=0}^\infty {1 \over \gamma ^{(2-D)\ n}}~e^{-\cos (\alpha )\ \gamma
^nx} ~\cos \left( \gamma ^nx\ \sin (\alpha )\right)~,  \label{genexpweier}
\ee
where 
\be
D=2-m=2-\frac{\ln \mu }{\ln \gamma }~.   \label{nbgnf}
\ee
The exponent $D$ turns out to be equal to the fractal dimension of the 
the graph of the Weierstrass function obtained for $\alpha=\pi/2$. Recall
that the fractal dimension quantifies the self-similarity properties of scale 
invariant geometrical objects. Note that $1<D<2$ as $1 < \mu < \gamma$
which is the condition of non-differentiability found by Hardy
\cite{Hardy} for the Weierstrass function. The graph of the Weierstrass function
is thus more than a line but less than a plane. For $\alpha < \pi/2$, $f(x)$ is smooth
and non-fractal ($D=1$) and its graph has the complexity of the line. 
Actually, there are several fractal dimensions. It is known that the
box counting (capacity, entropic, fractal, Minkowski) dimension and the packing
dimensions of the Weierstrass function are all equal to $D$ \cite{Kaplan} 
given by (\ref{nbgnf}) for $\alpha=\pi/2$. It is conjectured but not proved that
the Hausdorff fractal dimension of the graph of the Weierstrass
function obtained for $\alpha=\pi/2$ is also equal to $D$ given by (\ref{nbgnf}). 
It is known that the Hausdorff
dimension of the graph of $f(x)$ does not exceed $D$ but there is no satisfactory
condition to estimate its lower bound \cite{HuLau}.

Figure 2 shows the function (\ref{genexpweier}) for $\alpha = \pi/2=1.5708$ (pure 
Weierstrass function: panel a), $\alpha=0.993 \pi/2=1.56$ (panel b), 
$\alpha = 0.9 \pi/2 = 1.414$ (panel c) and
and $\alpha = 0$ (panel d). 

The Mellin transform of $g(x)$ defined by (\ref{exbifreg}) for
$-1 <$ Re$[s]=-m <0$ (which is the interval of interest, as seen
from (\ref{expgenfrac})) is \cite{tablesintegrals}
\be
\widehat{g}(s)= \Gamma (s) \cos (\alpha \ s) - {1 \over s}~, \label{mellingsss}
\ee 
where $\Gamma (s)$ is the Gamma function reducing to $\Gamma (s) =(s-1)!$
for integer arguments $s$. The additional term $-1/s$ disappears for 
$0<$Re$[s]$. For values of the exponent $m$ larger
than $1$, i.e., Re$[s]<-1$, additional correction terms should be 
added to (\ref{mellingsss}) \cite{tablesintegrals}. These
additional terms only contribute to the power law dependence of the amplitudes $A_n$
and not to the exponential. This problem is absent when the cosine
in the definition of $g(x)$ is replaced by the sine function.

As we shall discuss below, the modification of $g(x)$ into
the modified function 
\be
g_M(x)=e^{-\cos (\alpha )~x}~ \cos \left( x\ \sin (\alpha)\right) ~~-~1 ~
\label{modmamnnel}
\ee
gives $\widehat{g}(s)= \Gamma (s) \cos (\alpha \ s)$ without the correction
$-1/s$ for $-1 <$ Re$[s]=-m <0$ and leads to the so-called
Mandelbrot-Weierstrass function. 
Similar ``counter-term'' should be introduced for stretched exponential and in
similar cases. They do not bring any extra contributions to the Mellin
transform.

The regular part $f_r(x)$ defined by (\ref{reuughnla}) of $f(x)$ 
defined in (\ref{gensolsumsinreg}) 
corresponding to $g(x)$ defined by (\ref{exbifreg})  is
generated by the poles of $\Gamma (s)$, located at $s=-n$, $n=0,1...$, since
$\Gamma (s)$ is analytic on the whole complex plane excluding these simple poles
\cite{Lebedev}.
Using
the expression Re$s_{s=-n}\ \Gamma (s)=\frac{\left( -1\right) ^n}{n!}$, we 
obtain its explicit form (\ref{reuughnla})
with
\be 
B(n)=\frac{\left( -1\right) ^n}{n!}\frac \mu {\mu -\gamma ^n}\cos (\alpha n)~. 
\ee
Note the particularly simple expression of the first term $B_0= \frac \mu {\mu -1}$.
For $|x| \ll 1$, this constant term provides the only non-negligible contribution
of the regular part $f_r(x)$ to $f(x)$, whose behavior is completely controlled
by the sum $f_s(x)$ of singular power laws.

The amplitudes $A_n$ defined by (\ref{amplgensing}) corresponding 
to $g(x)$ defined by (\ref{exbifreg}) are
\be
A_n(\alpha)=\frac{\Gamma (s_n)\cos (\alpha \ s_n)}{\ln \gamma }~.  \label{amplifdhsk}
\ee
The singular part $f_s(x)$, which is defined by (\ref{sinsumgen})
where the exponents $s_n$ are given by (\ref{expgenfrac}), 
satisfies the exact scale-invariance equation (\ref{oneada}).

The asymptotic behavior of the amplitudes $A_n$ given by (\ref{amplifdhsk}) is
\be
A_n(\alpha )\sim {e^{\alpha m} \over n^{m+{1 \over 2}}}~ e^{-\omega \ n\ \left({\pi \over 2}
-\alpha \right)}~ e^{i~\omega ~n~\ln (\omega ~n)} ,\quad n\rightarrow \infty ~,
\label{asympAnkf}
\ee
with $m=\frac{\ln \mu }{\ln \gamma }$. The angular log-frequency $\omega$ is defined by
(\ref{guroww}).
In order to obtain (\ref{asympAnkf}), we have used the asymptotic dependence of 
the $\Gamma-$function asymptote for complex $z$ \cite{Lebedev}
\be
\Gamma (z) \simeq e^{ (z-1/2)\ln z-z}~,\quad \left| z\right| \gg 1. 
\ee
Expression (\ref{asympAnkf}) is of the form (\ref{classangener})
with $p=m+{1 \over 2}$, $\kappa =\omega \left({\pi \over 2} - \alpha\right)$ and
$\psi_n = \omega ~n~\ln (\omega ~n)$.

For $\alpha =0$,
\be
A_n(0)\sim {1 \over n^{m+{1 \over 2}}}~ e^{-{\pi \over 2} \omega \ n}~
e^{i~\omega ~n~\ln (\omega ~n)}~, \quad n\rightarrow \infty~. 
\ee
As we have shown above,
the fast exponential decay of $A_n(0)$ ensures the differentiability of $f(x)$
at all orders. Actually, the fast decay of $A_n(0)$ washes out any 
observable oscillatory structure from 
the function as seen on figure 2d. However, there are very tiny 
log-periodic oscillations of amplitude less than $5\cdot 10^{-7}$ (see table 1)
which are however unobservable at the scale of the plot of figure 2d.

For $\alpha =\pi /2$ (Weierstrass function), the exponential part disappears and
\be
A_n(\pi /2) \sim  {1 \over n^{m+{1 \over 2}}}~
e^{i~\omega ~n~\ln (\omega ~n)}~ ,\quad n\rightarrow \infty~.    \label{ghadfkq}
\ee
This situation corresponds to the case $p=m+{1 \over 2}$ and $\kappa =0$ and
$\psi_n = \omega ~n~\ln (\omega ~n)$ in
expression (\ref{classangener}) of 
the classification of section \ref{classecffr}.
The cancellation of the exponential term in $A_n$ is 
due to the very peculiar
compensation of the exponential decay of $\Gamma (s_n)$
by the exponential growth of $\cos (\alpha \ s_n)$ in (\ref{mellingsss}), which
occurs only for $\alpha = {\pi \over 2}$.

The original Weierstrass function (\ref{wiuelag}) is thus seen
as a very special ``critical'' or bifurcation point of the class of
``Weierstrass-type functions'' (\ref{glrfnaaglag}) with
(\ref{exbifreg}). The analogy goes further as
the expression (\ref{classangener}) for the amplitudes
$A_n$ has the same structure as the correlation function of a system of spins
where the order $n$ in the sum (\ref{sinsumgen}) plays the role of the distance $r$
between two spins. In this analogy, the ``correlation length'' is proportional 
to $1/\kappa  \sim \left({\pi \over 2} - \alpha\right)^{-1}$ and diverges at the
critical point $\alpha={\pi \over 2}$.

\subsection{Role of the phase: localization and delocalization of singularities}

The phases $\psi_n$ defined in (\ref{classangener}) play an essential role
in the construction of the self-affine nondifferentiable structure of the 
``Weierstrass-type functions''. 
To stress this fact, let us consider several cases using different phases
$\psi_n$ with the same absolute values $|A_n|$ of the amplitudes. This study
parallels in a sense that of Berry and Lewis \cite{Berry} and of Hunt \cite{Hunt}
but is distinct from it in a essential way
as the phases considered here decorate the amplitudes $A_n$
in (\ref{sinsumgen}) of the power series expansion, rather than the 
phases in the cosine in (\ref{wiuelag}). Actually, Berry and Lewis
study a slight modification of the Weierstrass function (\ref{wiuelag}) defined as
\be
f_{WM}=\sum_{n=0}^\infty b^n \left(1-\cos[a^{n} \pi x]\right)~,  \label{wiuafelag}
\ee
proposed by Mandelbrot \cite{Mandelbrot}, which has the property of directly 
satisfying the ``self-affine'' property (\ref{oneada}) with $\mu=1/b$ and $\gamma=1$.
As discussed above, the choice (\ref{modmamnnel}) for $g(x)$, which gives 
(\ref{wiuafelag}) up to a sign, has the advantage of getting rid of the $-1/s$ correction
in its Mellin transform (\ref{mellingsss}) which makes thus more apparent and direct its
self-similar properties.

Hunt \cite{Hunt} is 
able to show that, by replacing the argument $a^{n} \pi x$ of the cosine by
$a^{n} \pi x + \theta_n$ where $\theta_n$ are uncorrelated random phases, the
Hausdorff dimension of the phase-randomized Weierstrass function is $D=2-m$.

\subsubsection{Localization of singularities \label{locsinsec}}

Let us first study the case where $\psi_n$ is put equal
to $0$, i.e., we construct a phase-locked Weierstrass function as
\be
\overline{f_s(x)}=\sum_{n=0}^\infty \left| A_n(\pi /2)\right| \ x^{-s_n}~,  \label{gnlal}
\ee
i.e., by constructing the singular part as the sum over power laws with
amplitudes equal to the modulus of the amplitudes (\ref{amplifdhsk})
obtained for the Weierstrass function with $\alpha=\pi/2$, i.e., 
$|A_n(\pi/2)|=|\frac{\Gamma (s_n)\cos ((\pi/2) \ s_n)}{\ln \gamma }|$, but without
the phase. As a consequence, (\ref{ghadfkq}) is changed into
\be
|A_n(\pi /2)| =C  {1 \over n^{m+{1 \over 2}}}~, ~~~~ {\rm for}~ n \to \infty~,
\label{asyjlklwl}
\ee
where $C$ is a constant.

Figure 3 shows the function $\overline{f_s(x)}$ defined by (\ref{gnlal})
for $m=0.2$ (panel a) and $m=0.65$ (panel b). 
Rather than the familiar nondifferentiable self-affine corrugated structure
of the Weierstrass function, $\overline{f_s(x)}$ seems to be differentiable
everywhere except for a discrete infinity
of spikes at positions $x_u$, where $u$ is an integer running from $-\infty$
to $+\infty$, organized according to a geometric log-periodic structure.
 This discrete set of spikes decorates the leading singular behavior
$f(x) \sim  x^m$ for $x \to 0$ of the general solution
(\ref{gensolcrit}). Note that, in this case, the periodic function 
$P\left({\ln x \over \ln \gamma}\right)$ of the general solution
(\ref{gensolcrit}) is formed by the set of spikes geometrically converging to the 
origin.

The spikes seem to diverge for $m=0.2$ while they converge to a 
finite value for $m=0.65$,
as far as the numerical construction suggests. 
Appendix A examines some differentiability
properties of (\ref{gnlal}). 
Appendix B shows that 
the functional shapes of the spikes for $x \to x_u=1/\gamma^u$ with $u$ integer are
given by
\be
G_s(x) \sim {1 \over |x-x_u|^{{1 \over 2}-m}}~.
\ee
Thus, for $0 < m < 1/2$ (panel (a) of figure 3), the spikes correspond
to a divergence of $G_s(x)$ as $x \to x_u$. 
For $1/2 < m < 3/2$, $G_s(x)$ goes to a finite value as $x \to x_u$ but with an infinite
slope (since $0 < m-{1 \over 2} < 1$)
according to $G_s(x) \sim {\rm constant} - |x-x_u|^{m-{1 \over 2}}$. 

Another example of ``localization of singularities'' is provided by the function
\be
f_s(x)=\sum_{n=1}^\infty n^{-m-\frac 12}~ e^{i\ \omega ~\ln (\omega ~n)} \ x^{-s_n}~.   
\label{gnjlalkaaa}
\ee
Figure 4 shows this function $f_s(x)$ defined by (\ref{gnjlalkaaa}) with $m=0.2, \omega =7.7$.
One can observe
a log-periodic set of structures, each structure composed of log-periodic
oscillations converging to singular points beyond which damped oscillation can be observed.
Here, the phase
$\psi_n=\omega ~\ln (\omega ~n)$ is not varying fast enough with $n$ to scramble
the complex power laws $x^{-s_n}$, except at isolated points.

Figure 5 shows the graph of
\be
f_s(x)=\sum_{n=1}^\infty n^{-m-\frac 12}~ e^{i\ \omega ~ ~n} \ x^{-s_n}~,   
\label{gnjlaaalkaaa}
\ee
with $m=0.2, \omega =7.7$ and phases $\psi_n=\omega n$. Again, the phase are not
sufficiently random to make the function irregular, except at isolated points
where the constructive interference of the phases lead to the isolated
singularities. 

Note that both functions (\ref{gnjlalkaaa}) and (\ref{gnjlaaalkaaa}) can be 
analyzed with the method of Appendix B to obtain the functional form of the
singularities.

\subsubsection{Mixing phases}

In contrast to the previous examples where the phases $\psi_n$ are too regular, let us now take 
\be
\psi_n^{(0)} = \omega ~n~\ln (\omega ~n)     \label{ergophasefirst}
\ee
corresponding to the asymptotic
dependence (\ref{ghadfkq}) of the amplitudes $A_n$ of the Weierstrass function.
The phases (\ref{ergophasefirst}) are ergodic and mixing on the unit circle. 
The corresponding function is
\be
S(x)=\sum_{n=1}^\infty n^{-m-\frac 12}~ e^{i\ \omega ~n\ \ln (\omega ~n)} \ x^{-s_n}~,   
\label{gnjlalka}
\ee
which we call the ``log-periodic Weierstrass'' function to stress the fact
that it is constructed by summing log-periodic power laws $x^{-s_n}$
(see for instance expression (\ref{logperiodiccase}))
with amplitudes determined by the asymptotic behavior of the amplitudes
of the power expansion of the Weierstrass function itself. The exponents 
$s_n$ are again determined by (\ref{expgenfrac}) with $m$ given by (\ref{mvalue}).
By constructing (\ref{gnjlalka}),
we are stripping off the Weierstrass function of its regular part and of all 
features that are unrelated to its fundamental nondifferentiability and
self-affine properties. The definition of this ``log-periodic Weierstrass'' function
(\ref{gnjlalka}) and its many generalizations studied below examplifies
the novel construction developed here. $S(x)$ exhibits the same non-differentiability
as does the Weierstrass function. 
In all these cases, $S(x)$ is nondifferentiable since
$p=m+{1 \over 2} < {3 \over 2}$, in agreement with the conjecture of section 
\ref{classecffr}.

Consider the general case 
\be
S_i(x)=\sum_{n=1}^\infty n^{-m-\frac 12}\exp \left( i ~\psi_n^{(i)} \right) \ x^{-s_n}~.
\label{gnha;kdsa}
\ee


As other examples, let us now take
\be
\psi_n^{(1)}=\omega \ n^2~, \label{phase1}
\ee
and
\be
\psi_n^{(2)}= \omega~e^{n/\omega}  \label{phase2}
\ee
and form the corresponding sums for $i=1$ and $2$.
$\psi_n^{(0)}, \psi_n^{(1)}$ and $\psi_n^{(2)}$
cause similar irregular oscillations of $\cos[\psi_n^{(i)}]$ between $-1$ and $+1$
as a function of $n$, allowing for a
non-trivial and complex interactions of singularities whose amplitudes (most important
as we have seen)
exhibit a slow power-law decay. As the result $S_1(x)$ and $S_2(x)$ exhibit
very clear non-differentiable features shown in figures 9 and 10. 

Another very simple example of an ergodic phase is the
quadratic rotator with irrational rotation number $0 < R <1$:
\be
\psi_{n+1}^{(3)}=\psi_n^{(3)}+2\pi R n~.    \label{psofjjf}
\ee
The most irregular phase is obtained for $R=g=\frac{\sqrt{5}-1}{2}=0.61803398875...$
which is the golden mean whose main property is that it is the least-well
approximated by a rational number. 
The corresponding ``Golden-mean-log-periodic Weierstrass function'' $S_3^{(g)}(x)$
defined by (\ref{gnha;kdsa}) with (\ref{psofjjf}) is shown in figure 11. Other examples
with $R=\pi/4=0.785398163...$ and $R=1/e=0.367879441...$ lead to 
$S_3^{(\pi/4)}(x)$ and $S_3^{(1/e)}(x)$ shown in figures 12 and 13.
To each irrational number $R$ corresponds an interesting 
``Weierstrass-type functions'' whose delicately corrugated self-affine structure
is encoded in the number-theoretical properties of its corresponding irrational 
number $R$.

Note that if (\ref{psofjjf}) is changed into $\psi_{n+1}^{(3)}=\psi_n^{(3)}+2\pi R$,
the corresponding observable $f(x)$ becomes smooth almost everywhere except 
at isolated points (also organized according to a geometrical series as described
in appendix B). The ergodic but non-mixing properties of the linear rotation map 
does not scramble
the phases sufficiently to create non-differentiability.

\section{Illustration of the classification}

Different physical problems will be encoded by different regular parts $g(x)$
quantifying the impact on the observable $f(x)$ of the degrees of freedom
summed over between two successive magnifications with the ratio $\lambda$.
For a given physical problem, we perform
the Mellin transform of $g(x)$. Then, together with the renormalization
group structure (\ref{hghaqkq}), its formal solution (\ref{glrfnglag})
and its expansion (\ref{glrfnaaglag}) close to the leading critical point $x=0$
leading to ``Weierstrass-type functions'', the
classification (\ref{classangener}) of the previous section allows us
to characterize the possible non-differentiability and scaling properties
of the observable $f(x)$.

\subsection{Other examples of the $C^\infty$ differentiable family $\kappa>0$} 

For a general statistical mechanics model, the regular part $g(x)$ 
of the free energy has generally the form of the logarithm of a polynomial in $x$.
Factorizing the polynomial,
we do not lose generality by considering $g(x)$ given by
\be
g(x)=\ln(1+x)~,   \label{hgjgqqqq}
\ee
for which
\be
\hat{g}(s)={\pi\over s\sin s\pi}~.
\ee
The poles of $\hat{g}$ occur for $s=-n$, $n>0$, and contribute 
as already described only to the regular part of $f(x)$. 
Therefore, the Mellin transform of $f(x)$ is
\be
\widehat{f}(s)=\frac{\mu \gamma ^s}{\mu \gamma^s-1} ~\frac{\pi}{s \sin (s \pi )}~ .
\ee
The regular part, determined by the poles at $s=-n,\ n>0,$ reads
\be
f_r(x)=\sum_{m=1}^\infty B(m)\ x^m~,\quad \quad  \ B(m)=\frac{\left( -1\right) ^{m+1}}m
\frac \mu {\mu -\gamma ^m}~.   \label{jgnlqlqa}
\ee
Note that constant term is absent. The singular part is 
\be
f_s(x)=\sum_{n=0}^\infty A_n\ x^{-s_n}~,\quad \quad  A_n=\frac \pi {\ln \gamma
}\frac{1}{\sin (\pi ~ s_n)~ s_n}\ . 
\ee
The coefficients $A_n$ converge to zero extremely fast for large $n$
\be
A_n\sim \frac 1n\ e^{ -\pi \ \omega \ n}~ e^{-i\ \pi\ m}~,\ \quad \ \quad n\rightarrow \infty ~,
\label{jhnjgnlalq} 
\ee
where $\omega$ is given (\ref{guroww}). This corresponds to $p=1$, $\kappa=\pi \ \omega$
and $\psi_n=-\pi m$ in the general classification (\ref{classangener}). 

For the ``lorenzian'', 
\be
g(x)=\left( 1+x^2\right) ^{-1}\ ~,
\ee
the coefficients $A_n$ of the power law expansion (\ref{jgnlqlqa}) of
the singular part $f_s(x)$ are
\be
A_n=\frac \pi {2\ \ln \gamma }\frac 1{\sin (\pi /2\ s_n)\ }\ , 
\ee
with an asymptotic behavior given by
\be
A_n \sim \ e^{ -\frac \pi 2\omega n} ~e^{ i\ \frac{\pi}{2} m} ,\ \quad \ \quad n\rightarrow \infty~. 
\ee
This corresponds to $p=0$, $\kappa=\frac \pi 2\omega$
and $\psi_n=\frac{\pi}{2} m$ in the general classification (\ref{classangener}).
The exponential decay rate $\frac \pi 2\omega$ in this Lorentzian case is half that for the
logarithm function (\ref{hgjgqqqq}). Both lead to $C^\infty$ differentiable
functions with extremely small amplitudes of log-periodic oscillations (see table 1).

For so-called stretched exponential functions 
\be
g(x)= e^{-x^h}~, \quad \quad h>0~,  \label{ngnlgqlqq}
\ee
we obtain
\be
A_n(h)=\frac{\Gamma \left( s_n/h\right) }{\ln (\gamma )\ h}~, 
\ee
and 
\be
A_n(h)\sim {1 \over n^{{m \over h}+\frac 12}}~ e^{ -\frac{\pi \ \omega }{2\ h}\ n}~
e^{ i~ {\omega n\over h}~ \ln (\omega n)}~,\quad \quad  n\rightarrow \infty~. 
\ee
Log-periodic stretched exponential function $f_s(x,h)\ $and all its
derivatives (on $x$) converge. This corresponds to $p={m \over h}+\frac 12$, 
$\kappa={\pi \over 2} ~ {\omega \over h}$
and $\psi_n={\omega n\over h}~ \ln (\omega n)$ in the general 
classification (\ref{classangener}). Since $\kappa>0$, the corresponding
singular function $f_s(x)$ is differentiable at all orders. However, a
limit of non-differentiability at isolated points
can be reached formally by taking the limit $h \to \infty$
for which $p \to \frac 12$, $\kappa \to 0^+$ and and $\psi_n \to 0$. Then, $f_s(x)$
exhibits the non-differentiability at points $x_u$ verifying
$x_u=1/\gamma^u$ defined in (\ref{xudefres}) studied in section \ref{locsinsec}.
This is shown in figures 14 with the dependence of $f_s(x,h)$ on the parameter $h$. 
As $h$ increases, $g(x)$ becomes more and more localized close to the origin and
$f_s(x)$ exhibits more and more pronounced steps. Formally, the limit $h \to \infty$
allows us to cross-over from the class $\kappa>0$ to the class $\kappa=0$.

\subsection{Other examples of the ``Weierstrass-type function'' class $\kappa=0$} 

We have noted above that a
non-differentiable function is everywhere oscillating and the length of arc between
any two points on the curve is infinite \cite{Singh}. Its regular generator 
$g(x)$ must thus contain oscillations or must exhibit at least 
compact support (so that it has a discrete Fourier series) in order for $f(x)$ 
to be non-differentiable or for some of its derivatives to be non-differentiable.
We illustrate this remark by several examples.

\subsubsection{Generalized periodic processes}

Let us consider the function generalizing the sinc-function by taking an arbitrary
real exponent $\delta$:
\be
g(t)=\frac{\sin (t)}{t^\delta }~. \label{fnnggkza}
\ee
The Weierstrass function is recovered for $\delta = 0$.
The coefficients $A_n$ of the expansion in power series of
the singular part can be obtained by a simple shift of $s$ in the expression
obtained for the Weierstrass function:
\be
A_n(\delta )=\frac{\Gamma (s_n-\delta )\sin \left( \frac 12\pi \ (s_n-\delta
)\right) }{\ln \gamma \ }~. 
\ee
For $m+\delta \leq 1$, the log-periodic
generalized sinc-function $f_s(x,\delta )$ as well as its
associated Weierstrass-type function are continuous but non-differentiable.

The asymptotic behavior of the coefficient $A_n$ is
\be
A_n(\delta )\sim n^{-m-\delta -1/2} e^{ -i\ \omega \ n\ \ln (\omega n)}~ ,
\ \quad n\rightarrow \infty . 
\ee
This corresponds to $p=m+\delta +1/2$, $\kappa=0$
and $\psi_n=\omega n \ln (\omega n)$ in the general classification (\ref{classangener}).
Figure 15 shows the generalized sinc-function $f_s(x,\delta )$ for 
different values of $\delta$ with for $\delta =-0.1$ (solid line), 
$\delta=0$ (dashed line) and $\delta=0.1$ (dotted line)
for $m=0.4$ and $\omega =7.7$.

Another interesting case is the sine integral, $g(x)=Si(x)\equiv 
\int_0^x dv~{\sin v \over v}$. 
The coefficients $A_n$ of the power law series of the singular
part are given by
\be
A_n=-\frac{\Gamma (s_n)\sin (\frac 12\pi \ s_n)}{\ln \gamma \ s_n}~, 
\ee
with asymptotics 
\be
A_n\sim n^{-m-3/2}~e^{ -i\ \omega \ n\ \ln (\omega n) + {\pi \over 2}m}~,\  
\quad n\rightarrow \infty ~,
\ee
corresponding to $p=m+3/2$, $\kappa=0$
and $\psi_n=\omega n \ln (\omega n)+{\pi \over 2}m$ 
in the general classification (\ref{classangener}). 
All three functions $f(x)$, $f_s(x)$ and $S(x)$ defined by (\ref{gnjlalka})
have a continuous but non-differentiable first derivative for $m<1$.
However, the delicate log-periodic corrugations are enhanced in the graph of $S(x)$.

\subsubsection{Localized processes}

Let us now study functions $g(x)$ with compact support such as 
\be
g(x)=(1-x^h)^{\nu -1}~, \quad 0\leq x\leq 1~,\quad g(x)=0 ~~~{\rm for ~} x>1~~~
{\rm with}~ h\geq 1 ~~~{\rm and}~~\nu \geq 2~.   \label{gngmmcdaq}
\ee
The coefficients $A_n$ of the power series expansion of the singular part $F_s(x)$
are
\be
A_n(\nu ,h)=\frac{B\left( \nu ,s_n/h\right) }{\ln (\gamma )\ h} ~,
\ee
where $B(x,y)=\frac{\Gamma (x)\Gamma (y)}{\Gamma (x+y)}$ is the beta-function.
Figure 17 shows the
function $f(x)$ obtained from the direct sum (\ref{glrfnaaglag}). Figure 18
shows how the shape of the log-periodical structures steepen
with increasing $h$, as the function $g(x)$ evolves from a half-$\cap$ shape to a
the plateau $g(0 <x<1)=1$ and $0$ otherwise. The log-periodic
geometrical series of plateaux and steps shown in figure 18 is reminiscent
of the structures found for rupture \cite{Anifrani,canonical,critrup} and 
earthquakes \cite{SorSam,Salsamsor} precursors.

The asymptotic behavior of $A_n(\nu)$ is 
\be
A_n(\nu ,h)\sim \frac{e^{i \pi}}{n^\nu}~,\ ~~~{\rm with}~\nu \geq 2, \quad n\rightarrow \infty ~.   
\label{tjgnlgq}
\ee
corresponding to $p=\nu$, $\kappa=0$
and $\psi_n=\pi$ in the general classification (\ref{classangener}).

As the phases $\psi_n=\pi$ are constant and their contribution can
be factorized, the function $f_s(x,\nu)$ has a 
behavior similar to the function (\ref{gnlal}) analyzed in section \ref{locsinsec}.
In particular, we recover the fact that the points $x_u=1/\gamma^u$
given by (\ref{xudefres}) make the imaginary contribution of $x^{-s_n}$ 
vanish. As a consequence, they are the most singular points. An analysis
similar to that presented in section \ref{locsinsec} can be performed.

Another example corresponds to the logarithmic function 
\be
g(x)=\ln (1-x)~,~~~~~~~~~~0<x<1~~~~~{\rm and}~~~g(x) =0~~{\rm for}~~ x>1~,
\label{bgpgfqqf}
\ee
with compact support.
Figure 19 shows the corresponding Weierstrass function ($m=0.5, \omega =7.7, N=47$). 
The coefficients in the power expansion are given by
\be
A_n=-\frac 1{s_n}\left[ \Psi (1+s_n)-\Psi (1)\right] ~,
\ee
where $\Psi$ is the logarithmic derivative of the Gamma function. 
 
The asymptotic behavior of the coefficients $A_n$ is 
\be
A_n \sim \frac{\ln (n)}n\exp \left[ i\left( -~\arctan \left( \frac{2\ln
(\omega n)}\pi \right) +\pi \right) \right] ,\ \ \left( n\rightarrow \infty
\right) ~,   \label{gnhnllsls}
\ee
corresponding to $p=1$, $\kappa=0$
and $\psi_n=\arctan \left( \frac{2\ln (\omega n)}\pi \right)$ 
in the general classification (\ref{classangener}).
The logarithmic ``correction'' to the power law $1/n$ comes from the singularity at
$x=1$. This example illustrates a possible cause for a deviation from the
classification (\ref{classangener}).
Such modification however does not change the qualitative picture as they correspond
to the next sub-dominant correction to the power law contribution to $A_n$.

The relative amplitudes of the two first power law terms are given by
$\left| \frac{A_{n=1}}{A_{n=0}}\right| =0.143,\quad \left| \frac{A_{n=2}}{A_{n=0}}
\right| =0.086,\quad (m=0.5,\omega =7.7)$.

For the non-singular compact logarithmic regular part
\be
g(x)=\ln (1+x)~,~~~~~~~~~~0<x<1~~~~~{\rm and}~~~g(x) =0~~{\rm for}~~x>1~,
\label{nhgnnfnmbv}
\ee
figure 21 shows the corresponding Weierstrass-type function $f(x)$. The coefficients
in the power series expansion are 
\be
A_n=\frac 1{s_n}\left[ \log (2)-\frac 12\Psi (1+\frac{s_n}2)+\frac 12\Psi
(\frac 12+\frac{s_n}2)\right] ~,
\ee
where  $\Psi$ is again the logarithmic derivative of the Gamma function.
The asymptotic behavior of the coefficients $A_n$ is 
\be
A_n\sim \frac 1n\exp \left[ i\left( \arctan \left( \frac{\omega n}m\right)
+\pi \right) \right] ,\ \ \left( n\rightarrow \infty \right)~,
\ee
corresponding to $p=1$, $\kappa=0$
and $\psi_n=\arctan \left( \frac{\omega n}{m}\right) +\pi$ 
in the general classification (\ref{classangener}).
The relative amplitudes of the two first power law terms are given by
$\left| \frac{A_{n=1}}{A_{n=0}}\right| =0.016,\quad \left| \frac{A_{n=2}}{A_{n=0}}%
\right| =7.738~10^{-3}$ for $m=0.5,\omega =7.7$. Thus, even if the asymptotic
decay is almost the same as for (\ref{bgpgfqqf}) up to the logarithmic correction,
the logperiodic amplitudes of the leading terms are a factor of $10$ smaller.

\section{Discussion}

This paper has studied the solutions of the equation (\ref{glrfnaaglag}), which
can be understood as
a renormalization group equation with a single control
parameter or more generally as the Jackson $q$-integral describing 
discrete scale invariant (DSI) systems. We have put the
emphasis on the factors controlling the 
presence and amplitude of log-periodic corrections to the leading power law
solution. We have used the Mellin transform
to resum the formal series solution of the DSI equation into a power law
series and have presented a general classification within 
two classes:
\begin{enumerate}
\item systems with quasi-periodic ``regular part'' and/or with compact support
present strong log-periodic oscillatory amplitudes.

\item systems with non-periodic ``regular part'' with unbound support have
exceedingly small log-periodic oscillatory amplitudes and regular
smooth observables.
\end{enumerate}
In systems for which the renormalization group equation has been explicited, 
systems of the first class are associated with ``antiferromagnetic'' interactions.
Systems of the second class occur when the microscopic interactions are 
dominantly ``ferromagnetic''. These facts plus the observation that strong
log-periodic oscillations have been reported in out-of-equilibrium growth
processes, in rupture, in earthquakes and in finance lead us to propose
that strong logperiodicity is associated with the presence of interactions with
competing signs and long-range dipolar and multipolar form, favoring
different ``signs'' and modulated patterns. Dipolar and/or antiferromagnetic
interactions are well-known to lead to a rich behavior with striped patterns 
with alternating up and down spins (see \cite{dipolar} and references therein)
which are robust in the presence of disorder such as in the dipolar random-field
Ising model \cite{Natterman,Magni}. We conjecture that these spatial organization
translates into a quasi-periodic ``regular part'' (or with compact support) of the
renormalization group description.

The existence of ``antiferromagnetic'' or ``dipolar'' interactions leading to
coexistence of different ``signs'' and of modulated structures is documented
in rupture and finance. In rupture phenomena, cracking or local breaking leads
to a redistribution of stress according to the elastic ``dipolar'' Green function
which is non-monotonous and exhibit oscillations as a 
function of distance with local enhancement as well as screening alternating as 
a function of angular position. 
Similarly, an earthquake (corresponding technically to a ``double-couple'' force source
within the elastic crust) redistributes stress according to a clover-leaf pattern
with alternating $+$ and $-$ signs of the stress as a function of angular
position. This superficial analogy has been shown to be quite precise
in a mapping between the random fuse model and a dipolar magnet in which
the state of the network damage is associated with a metastable spin configuration
\cite{Barthelemy}.

Similarly, the complex price time series observed in financial stock markets
can be seen to result from the competition between ``value-searching'' investors
\cite{Boucont,Idesor}
who track the fundamental value of the stocks and ``trend-following'' traders
(see \cite{Techanajorg} and references therein)
who imitate the crowd thus developing herding behavior. 
Roughly speaking, the former (later) type of traders
sell (buy) when the price increases above the fundamental value and sell (buy) when
the reverse occurs. Thus, ``value-searching'' investors follow interactions similar to 
``antiferromagnetic'' coupling while the herding behavior resulting from the action of
``trend-following'' investors can be captured quantitatively by ``ferromagnetic''-type
interactions. 

We believe that the existence of non-monotonous interactions with competing signs
may be a fundamental mechanism of strong log-periodicity. Our analysis of
its impact on the ``regular part'' of the DSI equation suggests that the 
``antiferromagnetic'' and/or ``dipolar'' interactions are crucial ingredients at
the origin of fractal patterns in nature. 

However, determining the observables that take the place of the equilibrium free
energy for growth models which could be obtained recursively (via a
renormalisation group transformation) - or equivalently, the identification 
of the meaning of the
function $g(x)$, remains an unsolved problem. Correlatively, the mechanisms
in growth models
that select the log-period of the observed
log-periodic oscillations in the absence of a
hierarchical lattice have not been fully elucidated (see however \cite{Growth}
for a specific needle growth model and \cite{Sorreview} for a review). 
In a way, our paper has attempted to
provide an explanation of why the log-periodic oscillations may have a strong
amplitude, without explaining really their origin!
As a consequence, the program ahead of us
is i) to identify clearly the mechanism(s) underlying the emergence of DSI, ii) derive
the equivalent renormalization group equation. Only then, shall we be able to 
substantiate convincingly our conjectured mechanism in terms of ``anti-ferromagnetic''
interactions.

\vskip 0.5cm
{\bf Acknowledgments}: We are grateful to A. Erzan for a discussion
on Jackson's integral and for supplying the correponding references and to the referees'
suggestions that helped us improve the manuscript.

\pagebreak

\section*{Appendix A: Differentiability properties of the ``localization of singularities''}

Using the asymptotic expression (\ref{asyjlklwl}), we can write 
\be
{\rm Re}[\overline{f_s(x)}]= G(x)
+ x^m \sum_{n=n_r}^\infty  {1 \over n^{m+{1 \over 2}}} \ \cos\left(2 \pi n 
{\ln x \over \ln \gamma}\right)~,    \label{ghkanq}
\ee
where $G(x)= \sum_{n=0}^{n_r} \left| A_n(\pi /2)\right| \ x^{-s_n}$ is a regular 
function. ${\rm Re}[\overline{f_s(x)}]$ denotes the real part of $\overline{f_s(x)}$
and we have used (\ref{expgenfrac}). 
The second term of the r.h.s. of (\ref{ghkanq}), which can 
be called the singular part of ${\rm Re}[\overline{f_s(x)}]$
and is denoted $G_s(x)$, is a sum starting
at an index $n_r$ which is taken sufficiently large such that the asymptotic
expression (\ref{asyjlklwl}) holds to within any desired degree of accuracy.

The singular part $G_s(x)$ has the same analytical behavior as the function
\be
K_p(y) \equiv \sum_{n=1}^\infty  {1 \over n^{p}} \ \cos (n y) ~,~~~~p=m+{1 \over 2}~,
 \label{ngnqa}
\ee 
where $y=2\pi \ln x/\ln \gamma$. This function is a special case of
$K_{p,\{\psi_n\}}(y)$ defined by (\ref{jgnagn}) for $\psi_n=0$.

This function $K_p(y)$
has been studied in the literature for special cases. When the real
part of $p$ is larger than $1$, the sum is absolutely convergent for all $x$.
Restricting our attention to real exponents $p$, the series 
$K_p(y=\pi) = -\sum_{n=1}^\infty  {(-1)^{n+1} \over n^{p}}$, which
corresponds to $y=(2 \ell+1)\pi$ where $\ell$ is an arbitrary integer, is convergent
for all positive $p$ to $K_p(y=\pi) = (1-2^{1-p}) \zeta(p)$ \cite{Gradshtein}, where 
$\zeta(p)$ is the Riemann zeta function. Obviously, 
$K_p(y=2\pi)$ is infinite for $p<1$ and we show below that
$K_p(y \to 2\pi)$ has a power law singularity. For $p>1$, $K_p(y)$
can be expressed as 
\be
K_p(y) = {(2\pi)^p \over 4 \Gamma(p)}~ \sec\left({\pi p \over 2}\right)
~\left[ \zeta\left(1-p, {y \over 2\pi}\right) - \zeta \left(1-p, 1-{y \over 2\pi}\right)\right]~,
\ee
where $\zeta(s, \nu) = \sum_{k=0}^{+\infty} [\nu+k]^{-s}$ is the generalized
Riemann zeta function \cite{Prudnikov}.

For $p>0$ and except for the special value $y=0$ modulus $2\pi$, $K_p(y)$ is finite
and differentiable. To see this, let us consider rational values of $y/2\pi=r/q$ with
$q \geq 2$, where
$r/q$ is the irreducible representation of the rational $y/2\pi$. We can rearrange the
series in (\ref{ngnqa}) into $q$ sub-series as follows:
\ba
K_p(y)&=& {\rm Re}\left( \sum_{k=1}^\infty  {1 \over (kq)^{p}} 
+\sum_{k=1}^\infty  {e^{i 2\pi~ (q-1)(r/q)} \over (kq-1)^{p}}
+\sum_{k=1}^\infty  {e^{i 2\pi~ (q-2)(r/q)} \over (kq-2)^{p}} +... \right. \nonumber \\
&+& \left. \sum_{k=1}^\infty  {e^{i 2\pi~ 2(r/q)} \over (kq-q+2)^{p}} 
+\sum_{k=1}^\infty  {e^{i 2\pi ~(r/q)} \over (kq-q+1)^{p}} 
\right) =  {\rm Re} \sum_{j=1}^q \left( \sum_{k=0}^\infty {e^{i 2\pi ~j(r/q)} \over (kq+j)^{p}} 
\right) \nonumber \\
&=&  {\rm Re} \sum_{k=0}^\infty \left( \sum_{j=1}^q {e^{i 2\pi ~j(r/q)} \over (kq+j)^{p}} 
\right)
~, \label{ngnqaafda}
\ea
We expand $1/(kq+j)^{p}=(kq)^{-p}[1-{p \over kq} j + {p(p+1) \over 2} {j^2 \over (kq)^2}+...]$
and get
\be
K_p(y/2\pi=r/q) = {\rm Re} \sum_{k=0}^\infty {1 \over (kq)^p} \left( 
\sum_{j=1}^q e^{i 2\pi ~j(r/q)} - {p \over kq}
\sum_{j=1}^q j~e^{i 2\pi ~j(r/q)} + {p(p+1) \over 2(kq)^2}
\sum_{j=1}^q j^2~e^{i 2\pi ~j(r/q)} +...\right) ~.   \label{hghg}
\ee
Calling $w_q$ the $q$th root of $1$, i.e., $w_q= e^{i 2\pi/q}$, 
we have then $w_q+w_q^2+...+w_q^{q-1}+w_q^q=(1-w_q^q)/(1-w)=0$. Hence,
the first sum $\sum_{j=1}^q e^{i 2\pi ~j(r/q)}$ in (\ref{hghg})
is identically zero. The other sums are non zero and finite. We thus get
\be
K_p(y/2\pi=r/q) =  \sum_{k=0}^\infty {C_k(r,q) \over k^{p+1}}  ~.   \label{hghaag}
\ee
where 
\be
C_k(r,q)= {1 \over q^{p+1}}{\rm Re} \left[
\sum_{j=1}^q e^{i 2\pi ~j(r/q)} \left(-pj + {p(p+1) \over 2(kq)}j^2 -
{p(p+1)(p+2) \over 6(kq)^2}j^3 +...\right)\right]~
\ee
is bounded from above as $k \to +\infty$.
The expression (\ref{hghaag}) shows that $K_p(y/2\pi=r/q)$ is finite for any $p>0$.
Now, since rational numbers are dense among real numbers, i.e., any irrational
number can be approached arbitrarily close by a rational number, by the condition
of continuity, $K_p(y)$ is finite everywhere, except for $y/2\pi=r/q$ with $q=1$.
Differentiating the expression (\ref{ngnqa}) gives the series
$\sum_{n=1}^{\infty} {1 \over n^{p-1}} \ \sin (n y)$. By the same reasoning leading to
(\ref{hghaag}), this derivative is bounded from above by 
a constant times $\sum_{n=1}^{\infty} {1 \over n^{p}}$
which is convergent for $p>1$. This shows that $K_p(y)$ is differentiable for $p>1$, 
and thus ${\rm Re}[\overline{f_s(x)}]$ is differentiable for $m>1/2$. This approach
is not powerful enough however to treat the case $m<1/2$.

\pagebreak

\section*{APPENDIX B: 
Functional shape of the denumerable set of discrete singularities resulting from
the ``localization of singularities''}

We now examine the special case $y/2\pi=r/q$ with $q=1$.
From the expression (\ref{ghkanq}), the values $x_u$, which are such
that $\ln x_u / \ln \gamma$ is an integer $-u$, i.e.,
\be
x_u=1/\gamma^u~,   \label{xudefres}
\ee
make all the cosine terms
in the infinite sum in phase and equal to $1$. Thus, 
\be
G_s(x_u) = x^m \sum_{n=n_r}^\infty  {1 \over n^{m+{1 \over 2}}}~,    \label{ghaakanq}
\ee
which diverges for $m \leq 1/2$. At the border case $m=1/2$, the divergence is
logarithmic. Similarly, $dG_s/dx|_{x=x_u}$ diverges for $m<3/2$
as an additional power of $n$ is brought to each term in the sum by taking the
derivative. This and expression (\ref{xudefres}) explain the graphs of Figure 3. 

The functional shapes of the spikes for $x \to x_u$ can be determined as follows.
For $x \to x_u$, $\cos\left(2 \pi n  {\ln x \over \ln \gamma}\right)
= \cos\left({2 \pi n \epsilon \over \ln \gamma}\right) + {\cal O}(\epsilon^2)$, where 
$\epsilon \equiv (x-x_u)/x_u$ and ${\cal O}(\epsilon^2)$ represents a term proportional
to $\epsilon^2$. Let us now construct and compare $G_s(\epsilon)$ and 
$G_s(\lambda \epsilon)$, where $\lambda$ is an arbitrary number. Up to first order
in $\epsilon$, we have
\be
G_s(\lambda \epsilon) \approx
x_u^m \sum_{n=n_r}^\infty  {\cos\left({2 \pi n \lambda \epsilon \over \ln \gamma}\right)
\over n^{m+{1 \over 2}}}~.   \label{ghaakaaanq}
\ee
Posing $n'={\rm Int}(n \lambda)$, $G_s(\lambda \epsilon)$ can be rewritten
\be
G_s(\lambda \epsilon) \approx 
x_u^m \sum_{n'={\rm Int}[n_r \lambda]}^\infty  {\lambda^{m+{1 \over 2}} \over \lambda}~
{\cos\left({2 \pi n' \epsilon \over \ln \gamma}\right)
\over n'^{m+{1 \over 2}}} = x_u^m ~\lambda^{m-{1 \over 2}} ~
\sum_{n'={\rm Int}[n_r \lambda]}^\infty {\cos\left({2 \pi n' \epsilon \over \ln \gamma}\right)
\over n'^{m+{1 \over 2}}} ~.   \label{ghaafakaaanq}
\ee
Note the presence of the additional 
multiplicative term $\lambda^{m+{1 \over 2}} \over \lambda$ in the sum of (\ref{ghaafakaaanq}).
The numerator $\lambda^{m+{1 \over 2}}$ stems from replacing $n$ by $n'={\rm Int}(n \lambda)$
in $1/n^{m+{1 \over 2}}$. The other factor $1/\lambda$ is the ``Jacobian'' of the change
from $n$ to $n'={\rm Int}(n \lambda)$. Expression (\ref{ghaafakaaanq}) can then be rewritten
\be
G_s(\lambda \epsilon) \approx \lambda^{m-{1 \over 2}} ~G_s( \epsilon) + H_r(\epsilon)~,
\label{bgnhflaklaq}
\ee
where
\be
H(\epsilon) = \lambda^{m-{1 \over 2}}~
\sum_{n=n_r}^{{\rm Int}[n_r \lambda] -1} {\cos\left({2 \pi n' \epsilon \over \ln \gamma}\right)
\over n'^{m+{1 \over 2}}} 
\ee
is a regular function of $\epsilon$. The singular part of $G_s( \epsilon)$ 
is solution of 
$G_s(\lambda \epsilon) \approx \lambda^{m-{1 \over 2}} ~G_s( \epsilon)$, i.e., 
\be
G_s(\epsilon) \sim \epsilon^{m-{1 \over 2}}~.
\ee
This confirms that, for $0 < m < 1/2$ (panel (a) of figure 3), the spikes correspond
to a divergence of $G_s(x)$ as $x \to x_u$ according to $G_s(x) \sim 1/|x-x_u|^{{1 \over 2}-m}$.
For $1/2 < m < 3/2$, $G_s(x)$ goes to a finite value as $x \to x_u$ but with an infinite
slope (since $0 < m-{1 \over 2} < 1$)
according to $G_s(x) \sim {\rm constant} - |x-x_u|^{m-{1 \over 2}}$. The borderline case
$m=1/2$ can actually be summed exactly as $K_{p=1}(y)= -{1 \over 2} \ln 
\left(2[1-\cos y]\right)$ \cite{Gradshtein}. When 
$y \to 0$ modulo $2\pi$, $K_{p=1}(y)$ diverges
as $\ln {1 \over y}$ and thus $G_s(\epsilon)$ diverges as $G_s(\epsilon) 
\sim \ln \left|{x_u \over x-x_u}\right|$.

\pagebreak

\clearpage

\begin{table}[]
\begin{center}
\begin{tabular}{|c|c|c|c|c|c|} \hline
$g(x)$  & $p$ & $\kappa$ & $\psi_n$ & $\left|\frac{A_{n=1}}{A_{n=0}}\right|$ 
& $\left| \frac{A_{n=2}}{A_{n=0}} \right|$ \\ \hline
$\cos (x)$    & $m+1/2$ & $0$ & $\omega n\ln (\omega n)$ & $0.065$ &  $0.032$ \\ \hline
$\exp (-x)$    & $m+1/2$ & $\frac{\pi}{2} \omega$ & $\omega n\ln (\omega n)$ & $5.12~10^{-7}$ &  $1.432~10^{-12}$ \\ \hline
$\exp \left[ -c\ x\right] \cos \left( x\ s\right)^*$   
 & $m+1/2$ & $(\frac{\pi}{2}-\alpha) \omega$ & $\omega n\ln (\omega n)$ &  &   \\ \hline
$\left( 1+x^2\right)^{-1}$    & $0$ & $\frac{\pi}{2} \omega$ & $\frac{\pi}{2}m$ & $9.901~10^{-6}$ &  $4.414~10^{-11}$ \\ \hline
$\log (1+x)$    & $1$ & $\pi \omega$ & $-\pi m$ & $4.045~10^{-12}$ &  $\approx 0$ \\ \hline
$\exp (-x^h)$    & $m/h+1/2$ & $\frac{\pi}{2h} \omega$ & $\frac{\omega n}{h}\ln (\omega n)$ 
& $0.064 ~(h=50)$ &  $0.03~ (h=50)$ \\ \hline
    &  &  &   & $4.386~10^{-4} ~(h=2)$ &  $6.177~10^{-7} ~(h=2)$ \\ \hline
$\frac{\sin (x)}{x^\delta}$    & $m+\delta +1/2$ & $0$ & $-\omega n\ln (\omega n)$ 
& $0.044 ~(\delta=0.1)$ &  $0.021 ~(\delta=0.1)$ \\ \hline
   &  &  &   & $0.091 ~(\delta=-0.1)$ &  $0.049 ~(\delta=-0.1)$ \\ \hline
$Si(x)$    & $m+3/2$ & $0$ & $\omega n\ln (\omega n)$ & $4.199~10^{-3}$ &  $1.053~10^{-3}$ \\ \hline
$1-x^h$~~$0<x<1$ & $2$ & $0$ & $\pi$ & $0.064~ (h=50)$ &  
$0.031 ~(h=50)$ \\ \hline
 &  &  &   & $0.012 ~(h=2)$ &  $3.146~10^{-3} ~(h=2)$
\end{tabular}
\vspace{5mm}
\caption{\label{table1}  Synthesis of the different classes of Weierstrass-type
functions according to the general classification (\ref{classangener}) 
$A_n  \sim {1 \over n^p}~e^{- \kappa n}~e^{i \psi_n}$ of the expansion
(\ref{sinsumgen}) in terms of a series of power laws $x^{-s_n}$. The parameters
$p$, $\kappa  \geq 0$ and $\psi_n$ are determined by the form of $g(x)$ and the values of 
$\mu$ and $\gamma$. All numerical values given in this table correspond to
$m=0.5, \omega =7.7$ corresponding to $\gamma=2.26$ and $\mu=\sqrt{\gamma}=1.5$.
The last two columns quantify the amplitude of the log-periodic oscillations
with respect to the leading real power law.
$(*)$ $c=\cos \alpha$ and $s=\sin \alpha$.
}
\end{center}
\end{table}   

\clearpage

{\bf FIGURE CAPTIONS}
\vskip 0.5cm

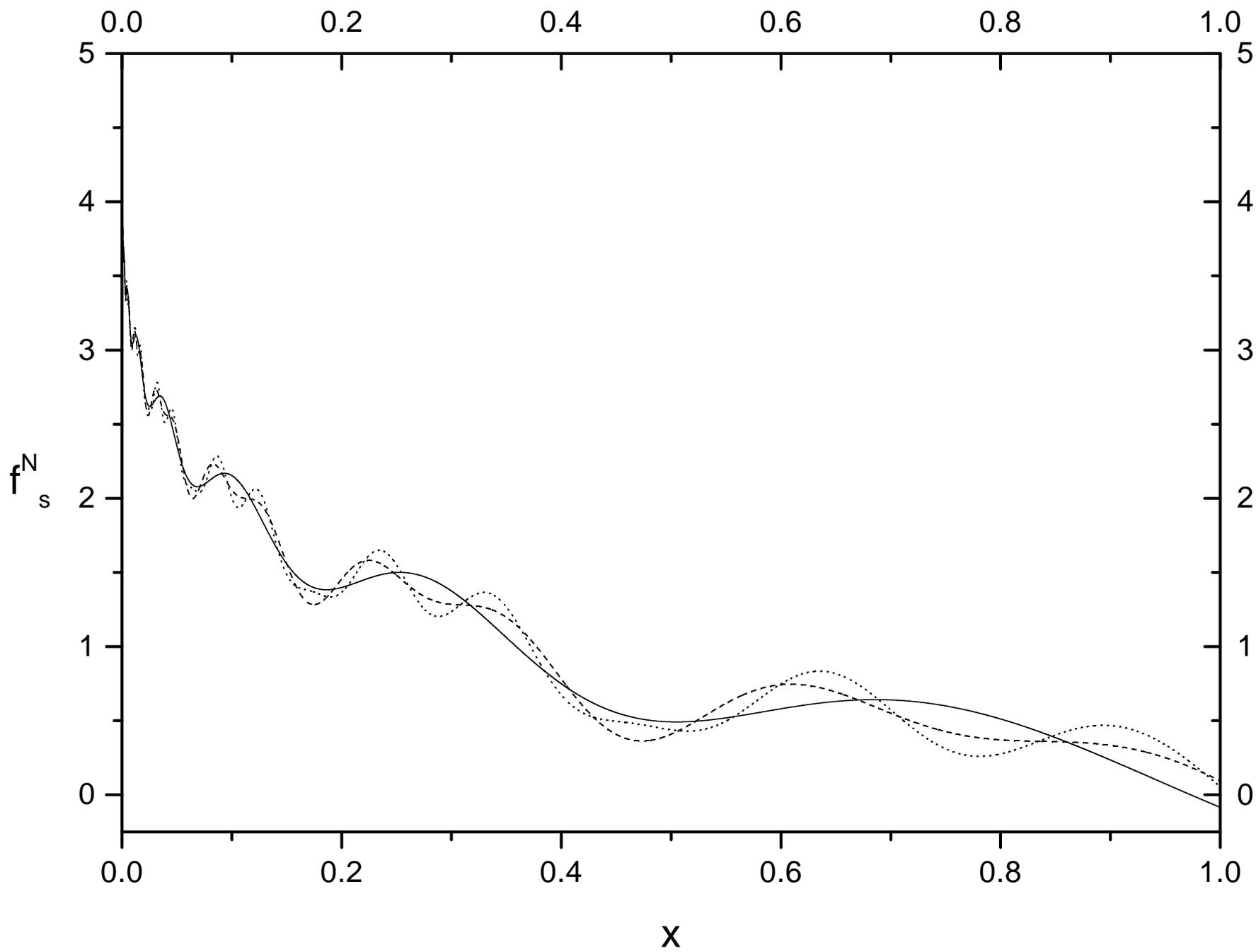
FIGURE 1: Power law expansion part $f_s$ given by (\ref{sinsumgen}) for the
Weierstrass function (\ref{wiuelag}), with 
$N=1$ (solid),  $N=2$ (dash), $N=3$ (dot) oscillatory terms, respectively.
Here, $m=0.25, \omega =6.3$ corresponding to $\gamma=2.7$ and $\mu=1.28$. As the 
number of complex exponents increases, the number of the oscillations increase. 

\vskip0.5cm
FIGURE 2: Quasi-Weierstrass function for 
(a) $\alpha =\frac \pi 2$, (b) $\alpha =0.993 \pi/2=1.56$, (c) $\alpha =0.9 \pi/2=1.414$
and (d) $\alpha =0$,  for $m=0.25, \omega =7.7$, using $N=32$ terms to estimate the sums
(\ref{genexpweier}). Increasing $N$ does not change the results
 
 \vskip0.5cm
FIGURE 3: Panels (a) and (b) show $f_s(x)$ defined by (\ref{gnlal})
with zero phase $\psi_n=0$, for $m=0.2$ and $m=0.65$ respectively, with the
same $\omega =7.7$, constructed by truncating the sum at the $N=29$ term.
The tiny regular oscillations result from the truncation to a finite $N$ and
slowly vanish when $N \to \infty$. They are thus spurious finite-size effects.
 
 \vskip0.5cm
FIGURE 4: Graph of $f_s(x)$ defined by (\ref{gnjlalkaaa})
with $\psi_n=\omega \ \ln (\omega n)$,  with $m=0.2, \omega =7.7$ and
using $N=1000$ terms in the sum.

 \vskip0.5cm
FIGURE 5: Graph of shows f$_s(x)$ defined by (\ref{gnjlaaalkaaa})
with $\psi_n=\omega n$, with $m=0.2, \omega =7.7$, with $N=1000$ terms in the sum.

 

 
 \vskip0.5cm
FIGURE 6: Graph of $S_1(x)$ defined by (\ref{gnha;kdsa}) for the phases
$\psi_n^{(1)}$ defined by (\ref{phase1} 
with $m=0.25, \omega =7.7,\ N=1000$.
 
 \vskip0.5cm
FIGURE 7 Graph of $S_2(x)$ defined by (\ref{gnha;kdsa}) for the phases
$\psi_n^{(2)}$ defined by (\ref{phase2} 
with $m=0.5, \omega =8,\ N=200$.

\vskip0.5cm
FIGURE 8: ``Golden-mean-log-periodic Weierstrass function'' $S_3^{(g)}(x)$
defined by (\ref{gnha;kdsa}) with (\ref{psofjjf}) for $m=0.5, \omega
=7.7,\ N=500$.

 \vskip0.5cm
FIGURE 9: ``$\pi/4$-log-periodic Weierstrass function'' 
$S_3^{(\pi/4)}$ defined by (\ref{gnha;kdsa}) with (\ref{psofjjf})
and $R=\pi/4$ for $m=0.5, \omega =7.7, N=500$.
 
 \vskip0.5cm
FIGURE 10: ``$e$-log-periodic Weierstrass function'' 
$S_3^{(e)}$ defined by (\ref{gnha;kdsa}) with (\ref{psofjjf})
and $R=e=2.718...$ for $m=0.5, \omega =7.7, N=500$.

 \vskip0.5cm
FIGURE 11: Singular part $f_s(x)$ of the Weierstrass-like function for
the regular function $g(x)$ equal to the stretched exponential (\ref{ngnlgqlqq})
for $h=5$ (solid line), $h=10$ (dashed line), $h=20$ (dotted line), $h=50$ (dashed-dotted
line) and $h=100$ (dashed-dot-dotted line), for
$m=0.4, \omega =7.7, N=22$.
 
 \vskip0.5cm
FIGURE 12: Singular part $f_s(x, \delta)$ of the Weierstrass-like function for
the regular function $g(x)$ given by (\ref{fnnggkza})
for $\delta =-0.1$ (solid line),  $\delta=0$ (dashed line) and $\delta=0.1$ (dotted line)
for $m=0.4,\omega =7.7,\ N=27)$.


 \vskip0.5cm
FIGURE 13: Weierstrass-type function $f(x)$ for compact $g(x)$ given by (\ref{gngmmcdaq})
with $\nu =2, h=2$ for $m=0.5, \omega =7.7,\ N=47)$.
 
 \vskip0.5cm
FIGURE 14: Evolution of the ``singular part'' $f_s(x)$ corresponding to
the compact regular part $g(x)$ (\ref{gngmmcdaq}) 
for $\nu =2$ with increasing abruptness of $g(x)$ quantified by the 
exponent $h$: $h=2$ (solid), $h=5$ (dash), $h=10$ (dot), $h=20$
(dash dot), $h=50$ (dash dot dot), $h=100$ (short dash),
for $m=0.5, \omega =7.7,\ N=47)$.

 \vskip0.5cm
FIGURE 15: Weierstrass-type function $f(x)$ corresponding to the regular part
$g(x)$ defined by (\ref{bgpgfqqf}) with compact support, with 
$m=0.5,\omega =7.7, N=47$.
 
 
 \vskip0.5cm
FIGURE 16: Weierstrass-type function $f(x)$ corresponding to the regular part
$g(x)$ defined by (\ref{nhgnnfnmbv}) with compact support, with
$m=0.5,\omega =7.7, N=47$.
 


\begin{thebibliography}{99}

\bibitem{Sorreview}  D. Sornette, Discrete Scale Invariance and Complex Dimensions,
Physics Reports 297, 239-270 (1998).

\bibitem{DIL} B. Derrida, C. Itzykson and J.M. Luck, 
Oscillatory critical amplitudes in hierarchical models, 
Commun. Math. Phys. 94, 115-132 (1984).

\bibitem{DLA} D. Sornette, A. Johansen,  A. Arn\'eodo, J.-F. Muzy and H. Saleur,
Complex fractal dimensions describe the internal hierarchical structure of DLA, 
Phys. Rev. Lett. 76, 251-254 (1996).

\bibitem{Growth} Y. Huang, G. Ouillon, H. Saleur and D. Sornette,
 Spontaneous generation of discrete scale invariance in growth models, 
 Physical Review E 55, N6, 6433-6447 (1997).

\bibitem{critrup} A. Johansen and D. Sornette,
Critical ruptures, Eur. Phys. J. B 18, 163-181 (2000).

\bibitem{SorSam} D. Sornette and C.G. Sammis,
Complex critical exponents from renormalization group theory of earthquakes:
Implications for earthquake predictions, J.Phys.I France 5, 607-619 (1995).

\bibitem{SJB}  D. Sornette, A. Johansen and J.-P. Bouchaud,
Stock market crashes, Precursors and Replicas, J.Phys.I France 6, 167-175  (1996).

\bibitem{JSL}  A. Johansen, D. Sornette and O. Ledoit, Predicting Financial
crashes using discrete scale invariance, Journal of Risk 1, 5-32 (1999).

\bibitem{Jacksonqder} F.H. Jackson, Q.J. Pure. Appl. Math 41, 193 (1910);
Q.J. Math Oxford Ser. 2, 1 (1951).

\bibitem{Erzan1} A. Erzan, Finite $q$-differences and the discrete renormalization 
group, Phys. Letts. A 225, 235-238 (1997).

\bibitem{ErzanEck} A. Erzan and J.-P. Eckmann, $q$-analysis of Fractal Sets,
Phys. Rev. Letts. 78, 3245-3248 (1997).

\bibitem{BEE}  B. Derrida, J-P Eckmann and A. Erzan, Renormalization groups
with periodic and aperiodic orbits, J. Phys. A, 16, 893-906 (1983).

\bibitem{McKay} S. R. McKay, A.N. Berker and S. Kirkpatrick, Spin-glass behavior
in frustrated Ising models with chaotic renormalization group trajectories,
Phys. Rev. Letts. 48, 767-770 (1982).

\bibitem{Paul} G. Paul, Coefficient scaling, Phys. Rev. E 59, 4847-4856 (1999).

\bibitem{Dubrulleetal} B. Dubrulle, F. Graner and D. Sornette, eds.,
Scale invariance and beyond, (EDP Sciences and Springer, Berlin, 1997).

\bibitem{mybook} D. Sornette, Critical Phenomena in Natural Sciences 
(Chaos, Fractals, Self-organization and Disorder: Concepts and Tools)
(Springer Series in Synergetics, 2000).

\bibitem{Goldenfeld} N. Goldenfeld, 
Lectures on phase transitions and the renormalization group
(Reading, Mass.: Addison-Wesley, Advanced Book Program, 1992).

\bibitem{Erzanderivation} A. Erzan, Phys. Lett. A 225, 235 (1997).

\bibitem{Weierstrass} K. Weierstrass, On Continuous Functions of a Real Argument that
do not have a Well-defined Differential Quotient, in Classics on
Fractals, Edited by G. A. Edgar (Addison-Wesley, Reading, Massachusets,
1993) p. 3-9.

\bibitem{Hardy} G.H. Hardy, Weierstrass's non-differentiable function, Trans. Amer. Math. Soc. 
17, 301-325 (1916).

\bibitem{Singh} A.N. Singh, The theory and construction of non-differentiable functions,
in Squaring the circle and other monographs (Chelsea Publishing Company, 1953).

\bibitem{Richardson}  L. F. Richardson, Atmospheric Diffusion on a Distance-Neighbour
Graph, Proc. Roy. Soc. London A 110, 709 (1926).

\bibitem{Shlesinger} B. D. Hughes, M.F. Shlesinger, and E. W. Montroll, 
Random walks with self-similar clusters, Proc. Natl. Acad.
Sci. U.S.A. 78, 3287-3291 (1981); 
E.W. Montroll and M. F. Shlesinger, in
Studies in Statistical Mechanics, edited by J. Lebowitz and E. Montroll
(North-Holland, Amsterdam, 1984), Vol. 11, p. 1;
J. Klafter, M>F> Shlesinger and G. Zumofen, Beyond Brownian motion, 
Physics Today 49, 33-39 (1996); M.F. Shlesinger, G.M. Zaslavsky and J. Klafter,
Strange kinetics, Nature 363, 31-37 (1993).

\bibitem{salsor} H. Saleur and D. Sornette,
Complex exponents and log-periodic corrections in frustrated systems, 
J.Phys.I France 6, 327-355 (1996).

\bibitem{lapidus} M.L. Lapidus and M. van Frankenuysen, Fractal geometry and
number theory (Birch\"auser, Boston, Basel, Berlin, 2000).

\bibitem{Sidorov} Yu.V. Sidorov, M. V. Fedoryuk and M. I. Shabunin, Lectures in Theory of
Complex Variable, Nauka, Moscow (1989), p.480.

\bibitem{Oberhett} F. Oberhettinger, Tables of Mellin Transforms, Springer-Verlag, N.Y.
(1974), p.275

\bibitem{Anifrani} J.-C. Anifrani, C. Le Floc'h, D. Sornette and B. Souillard,
Universal Log-periodic correction to renormalization group scaling for rupture stress
prediction from acoustic emissions, J.Phys.I France 5, 631-638 (1995).

\bibitem{canonical} A. Johansen and D. Sornette,
Evidence of discrete scale invariance by canonical averaging,
Int. J. Mod. Phys. C 9, 433-447 (1998).

\bibitem{Salsamsor} H. Saleur, C.G. Sammis and D. Sornette,
Discrete scale invariance, complex fractal dimensions and log-periodic
corrections in earthquakes, J. Geophys. Res. 101, 17661-17677 (1996).

\bibitem{Johsalsor} A. Johansen, H. Saleur and D. Sornette,
New Evidence of Earthquake Precursory Phenomena in the
17 Jan. 1995 Kobe Earthquake, Japan, Eur. Phys. J. B 15, 551-555 (2000).

\bibitem{JSNasdaq}  A. Johansen and D. Sornette, The Nasdaq crash of April 2000: Yet
another example of log-periodicity in a speculative bubble ending in a
crash, Eur. Phys. J. B 17, 319-328 (2000).

\bibitem{Klafteretal}  R. Metzler, J. Klafter and J. Jortner, Hierarchies and
logarithmic oscillations in the temporal relaxation patterns of proteins and
other complex systems, Proc. Natl. Acad. Sci. USA 96, 11085-11089 (1999).

\bibitem{HuLau} T.-Y. Hu and K.-S. Lau, Fractal dimensions and singularities of
the Weierstrass-type functions, Trans. Am. Math. Soc. 335, 649-665 (1993).

\bibitem{Titchmarsh} E.C. Titchmarsh, The Zeta-Function of Riemann, Cambridge University
Press, London, 1930, p. 104.  

\bibitem{Edwardszeta} H.M. Edwards, Riemann's zeta function, Academic Press, New York (1974). 

\bibitem{Kaplan} J.L. Kaplan, J. Mallet-Paret and J.A. Yorke, The Lyapunov dimension
of a nowhere differentiable attracting torus, Ergod. Th. Dynam. Sys. 4, 261-281 (1984).

\bibitem{tablesintegrals}  Tables of integral transforms, Edited by. A. Erde$^{^{\prime }}$%
lyi, McGraw-Hill, New York, 1954.

\bibitem{Lebedev}  N.N. Lebedev, Special functions and their applications,
Englewood Cliffs, N. J., Prentice-Hall, 1965, p.308.

\bibitem{Gradshtein} I. S. Gradshteyn and I.M. Ryzhik, 
Table of integrals, series, and products, 4th ed. prepared by Yu. 
V. Geronimus and M. Yu. Tseytlin,
Translated from the Russian by Scripta Technika, inc. Translation edited by
Alan Jeffrey (New York, Academic Press, 1965).

\bibitem{Berry}  M. V. Berry and Z. V. Lewis, On the Weierstrass-Mandelbrot
fractal function, Proc. Roy. Soc. A 370, 459-484 (1980).

\bibitem{Hunt} B.R. Hunt, The Hausdorff dimension of graphs of Weierstrass functions,
Proc. Am. Math. Soc 126, 791-800 (1998).

\bibitem{Mandelbrot} B. Mandelbrot, The fractal Geometry of Nature (Freeman, 
San Francisco, 1982).

\bibitem{Prudnikov} A.P. Prudnikov, Yu. A. Brychkov, O.I. Marichev,
Integrals and series, translated from the Russian by N.M. Queen (New York: Gordon and Breach
Science Publishers, 1986-1992).

\bibitem{dipolar} K. De'Bell, Maclsaac, A.B. and Whitehead, J.P.,
Dipolar effects in magnetic thin films and quasi-two-dimensional systems,
Rev. Mod. Phys. 72, 225-257 (2000).

\bibitem{Natterman} T. Natterman, Dipolar interaction in random-field systems,
J. Phys. A: Math. Gen. 21, L645-L649 (1988).

\bibitem{Magni} A. Magni, Hysteresis properties at zero temperature 
in the dipolar random-field Ising model, Phys. Rev. B 59, 985-990 (1999).

\bibitem{Barthelemy} M. Barth\'el\'emy, R. da Silveira and H. Orland, The
random fuse network as a dipolar magnet, cond-mat/0106012

\bibitem{Boucont} Bouchaud, J.-P. and R. Cont  (1998)
A Langevin approach to stock market fluctuations and crashes,
Eur. Phys. J. B 6, 543-550.

\bibitem{Idesor} K. Ide and D. Sornette,
Oscillatory Finite-Time Singularities in Finance, Population and Rupture, 
preprint  (e-print at http://arXiv.org/abs/cond-mat/0106047)

\bibitem{Techanajorg} Andersen, J.V., S. Gluzman and D. Sornette (2000)
Fundamental Framework for Technical Analysis, European Physical 
Journal B {\bf 14}, 579-601.

\end{thebibliography}
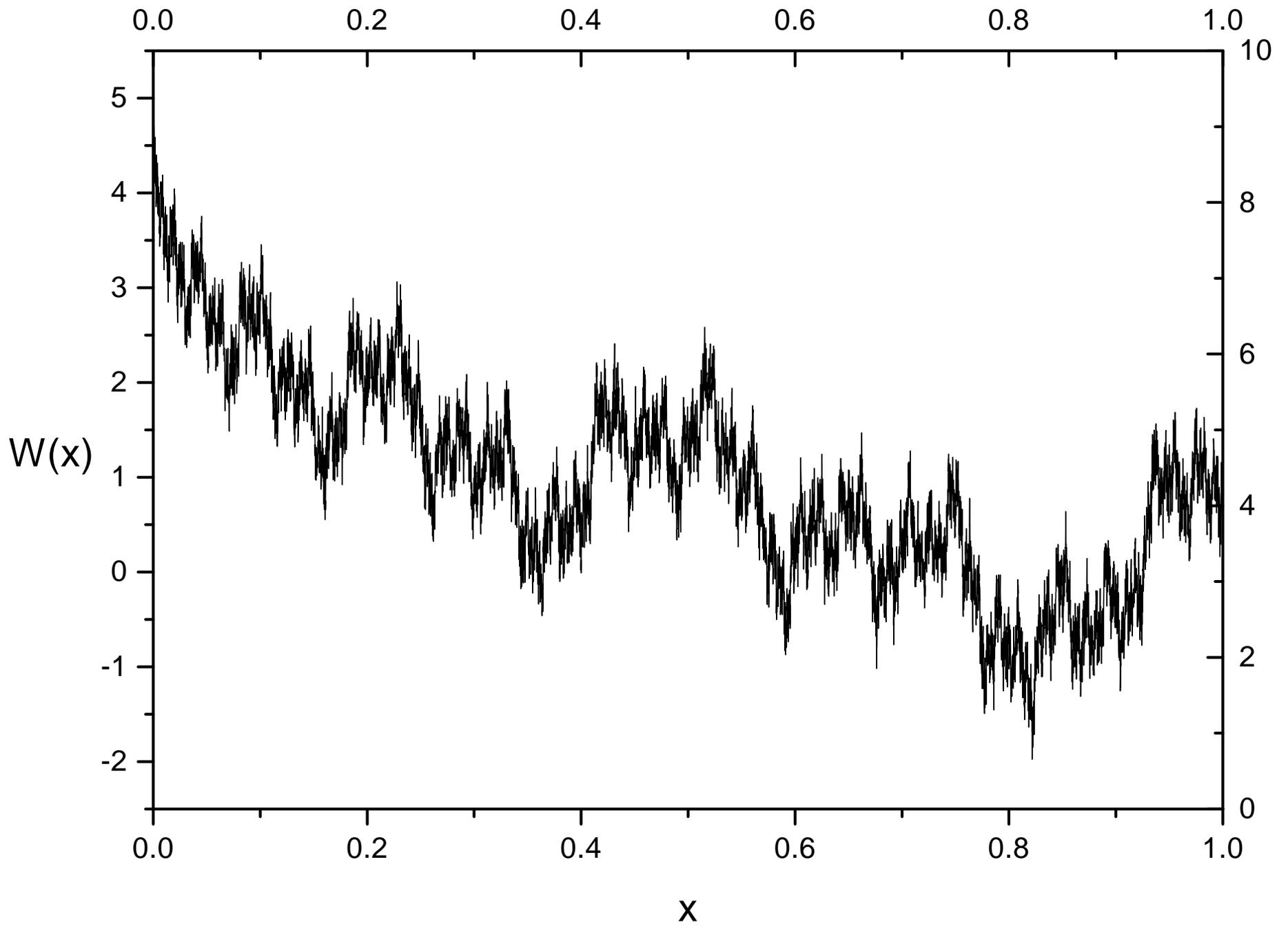
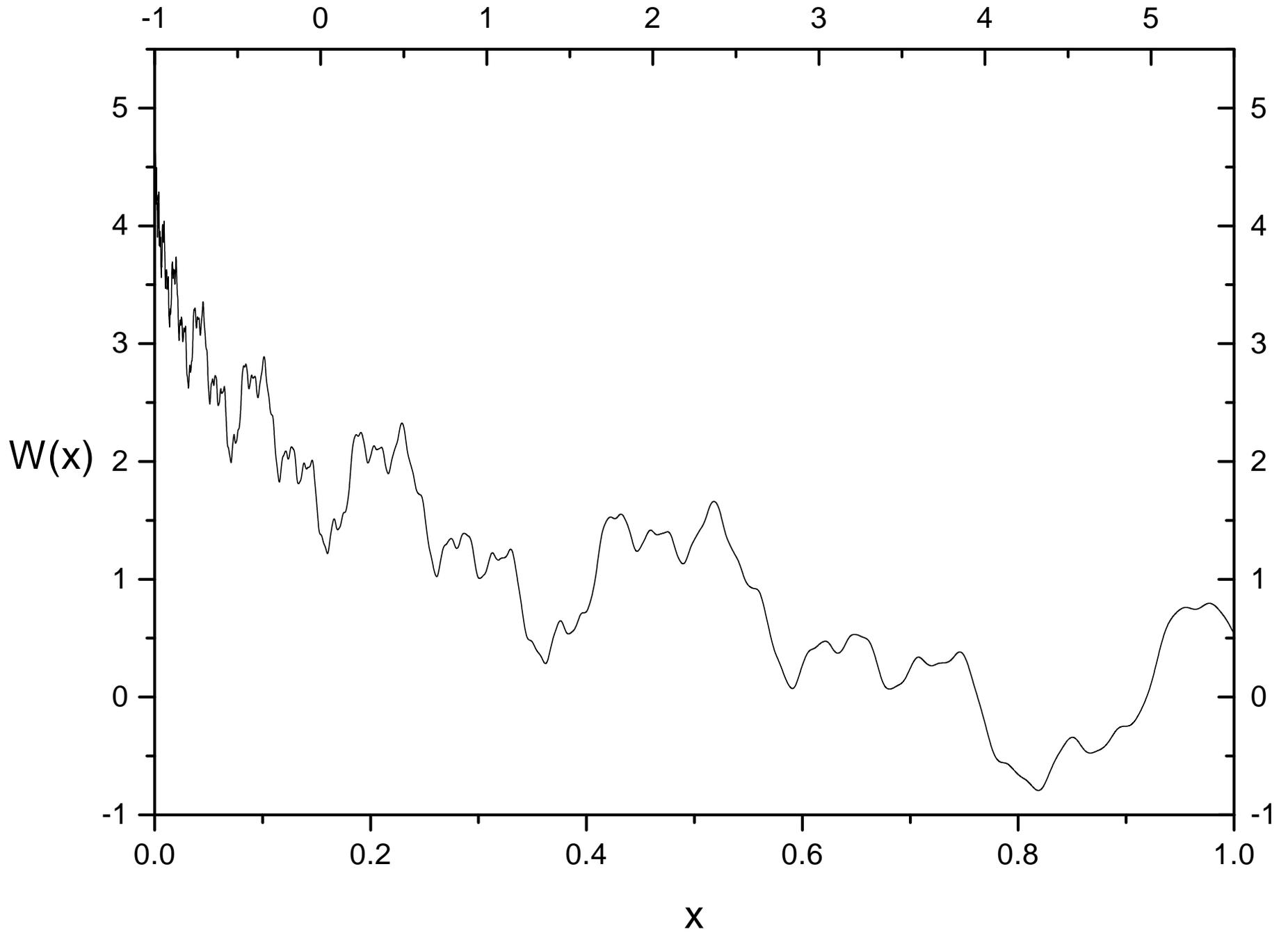
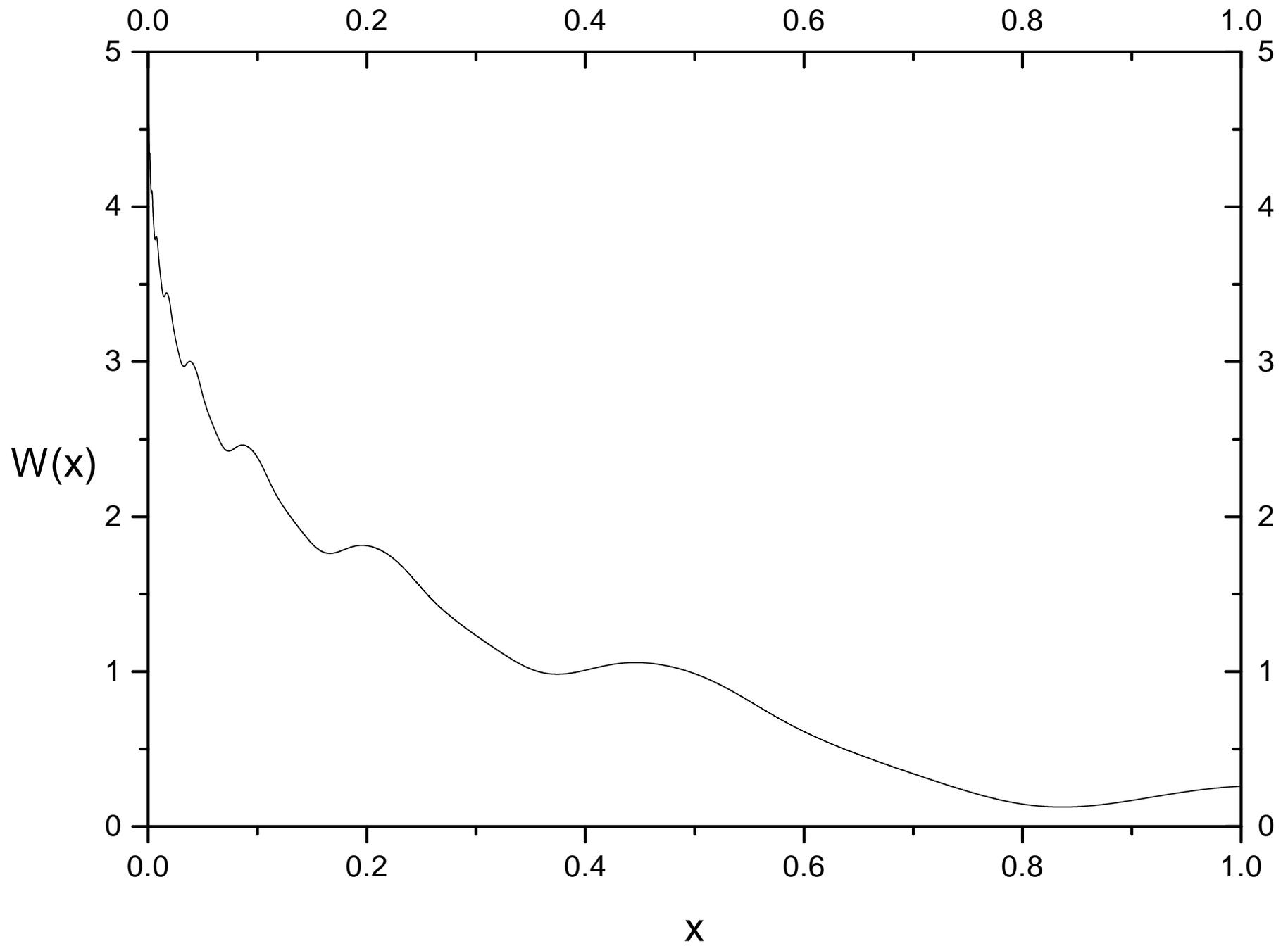
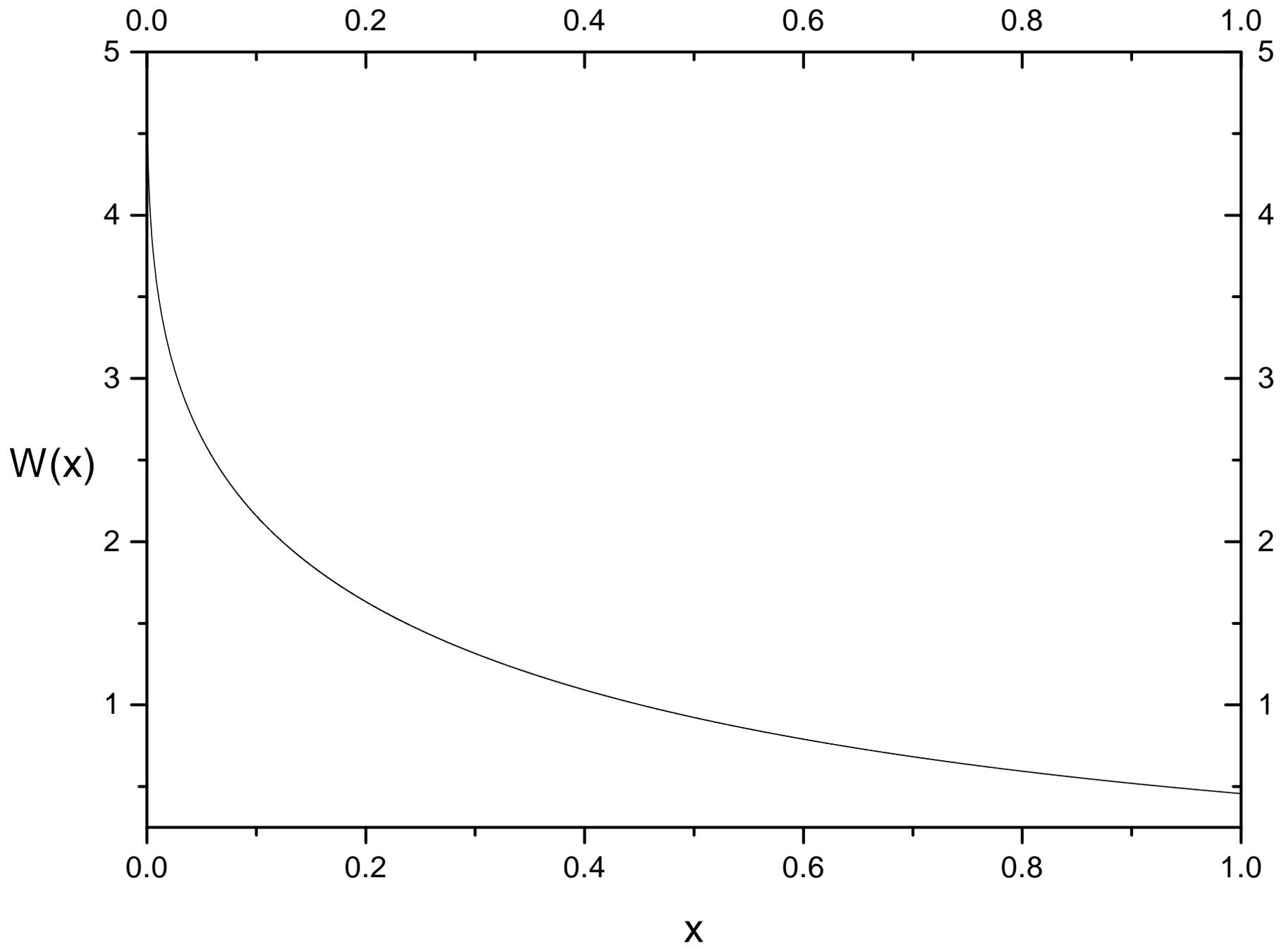
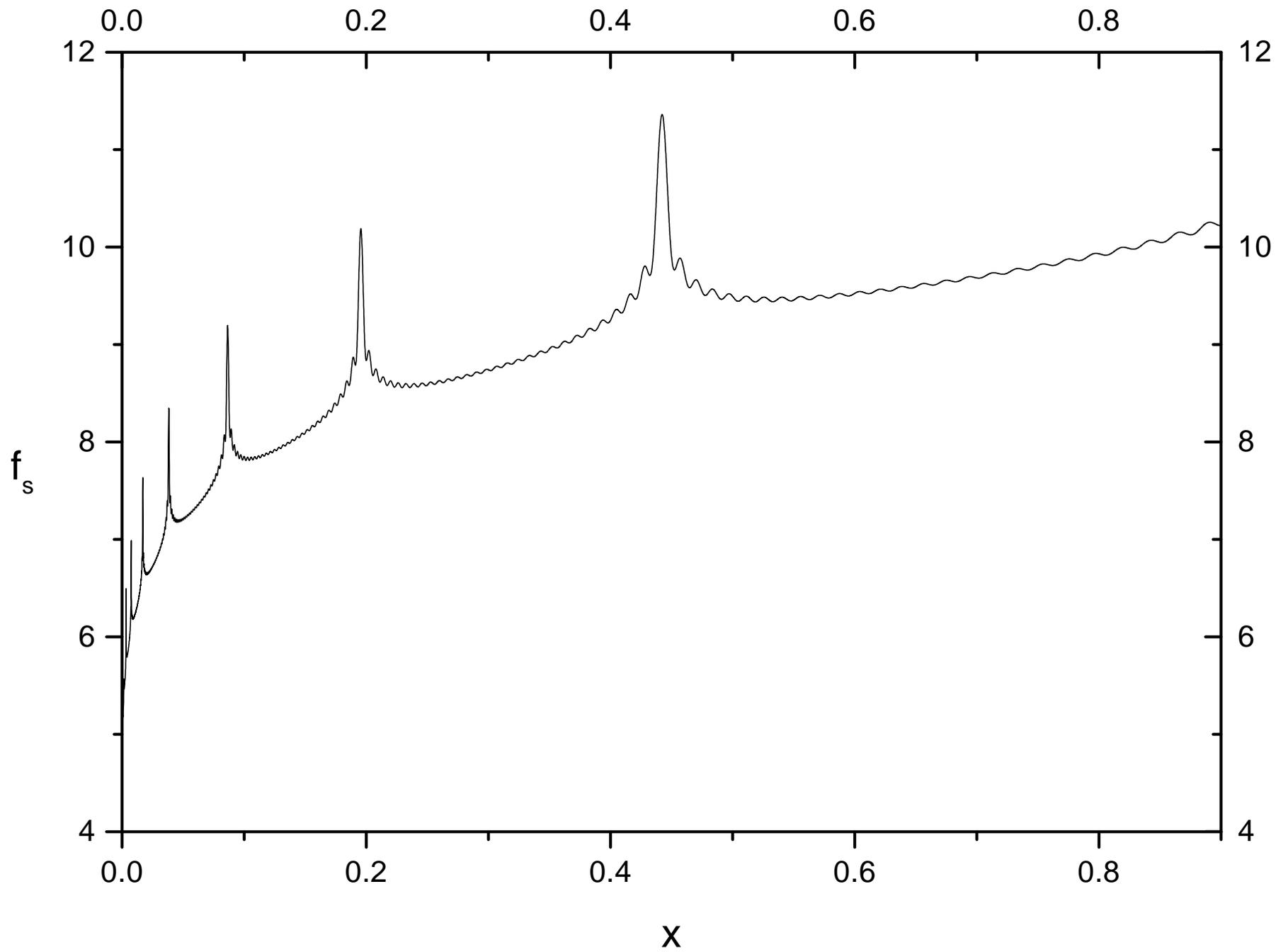
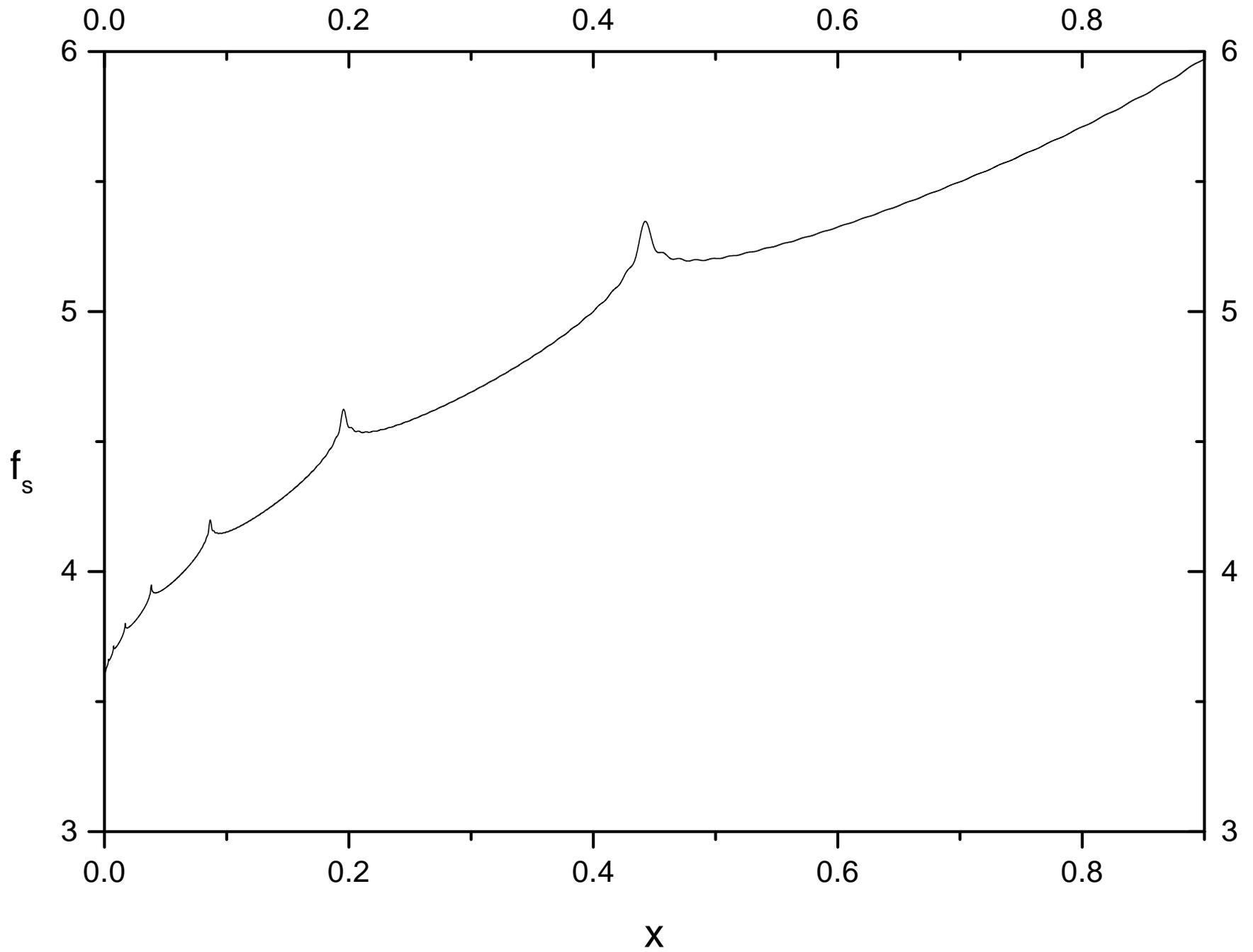
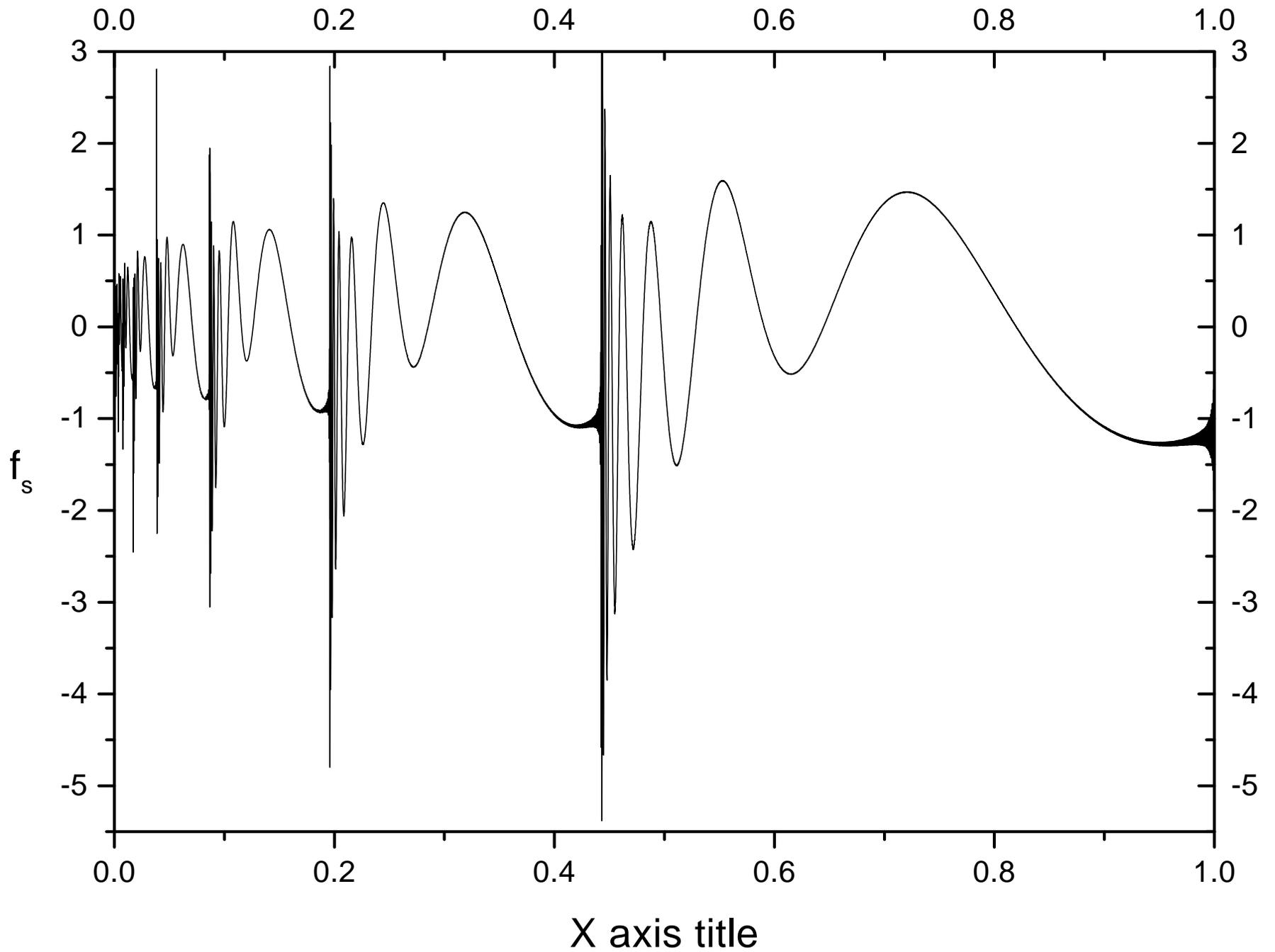
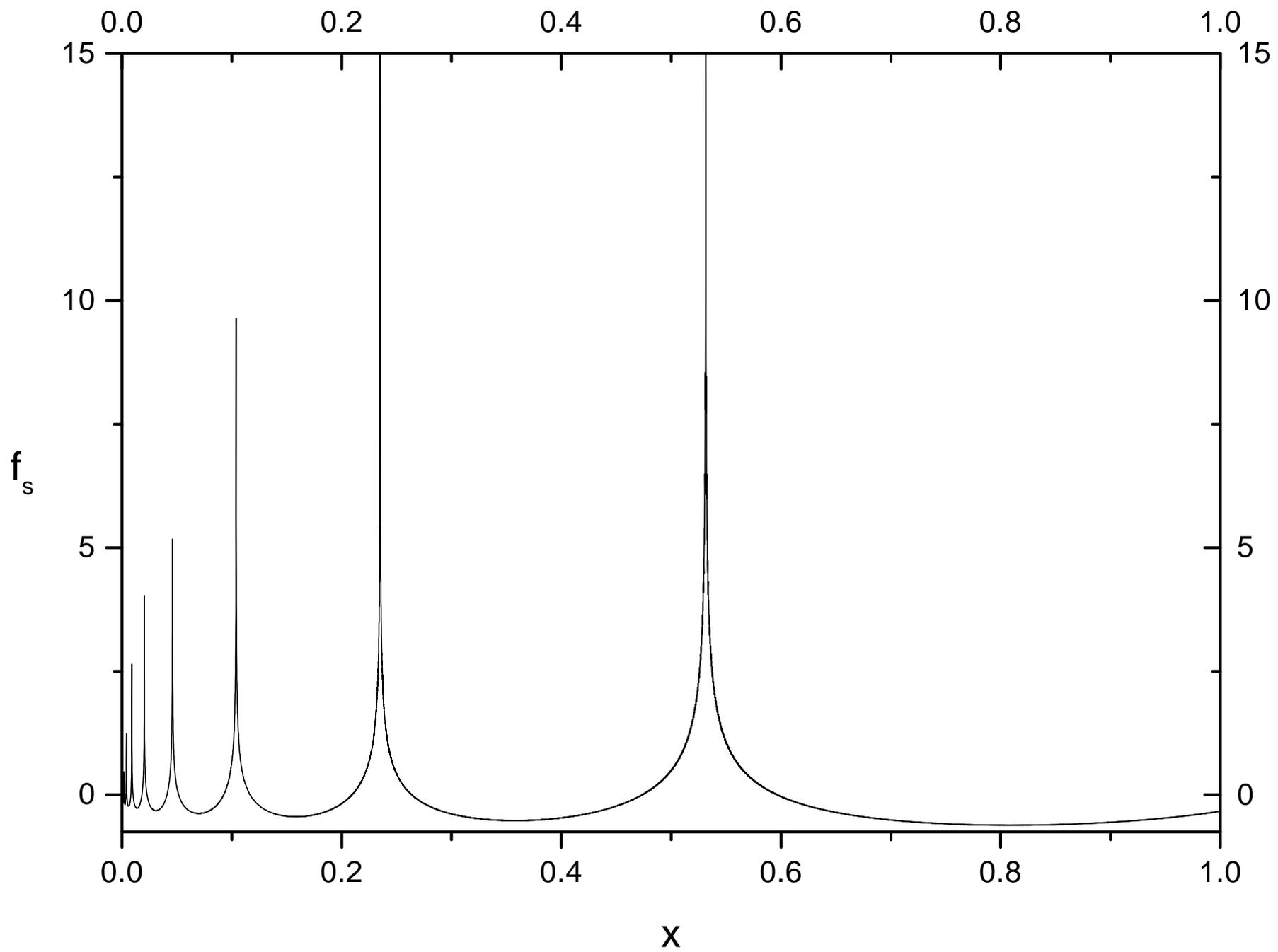
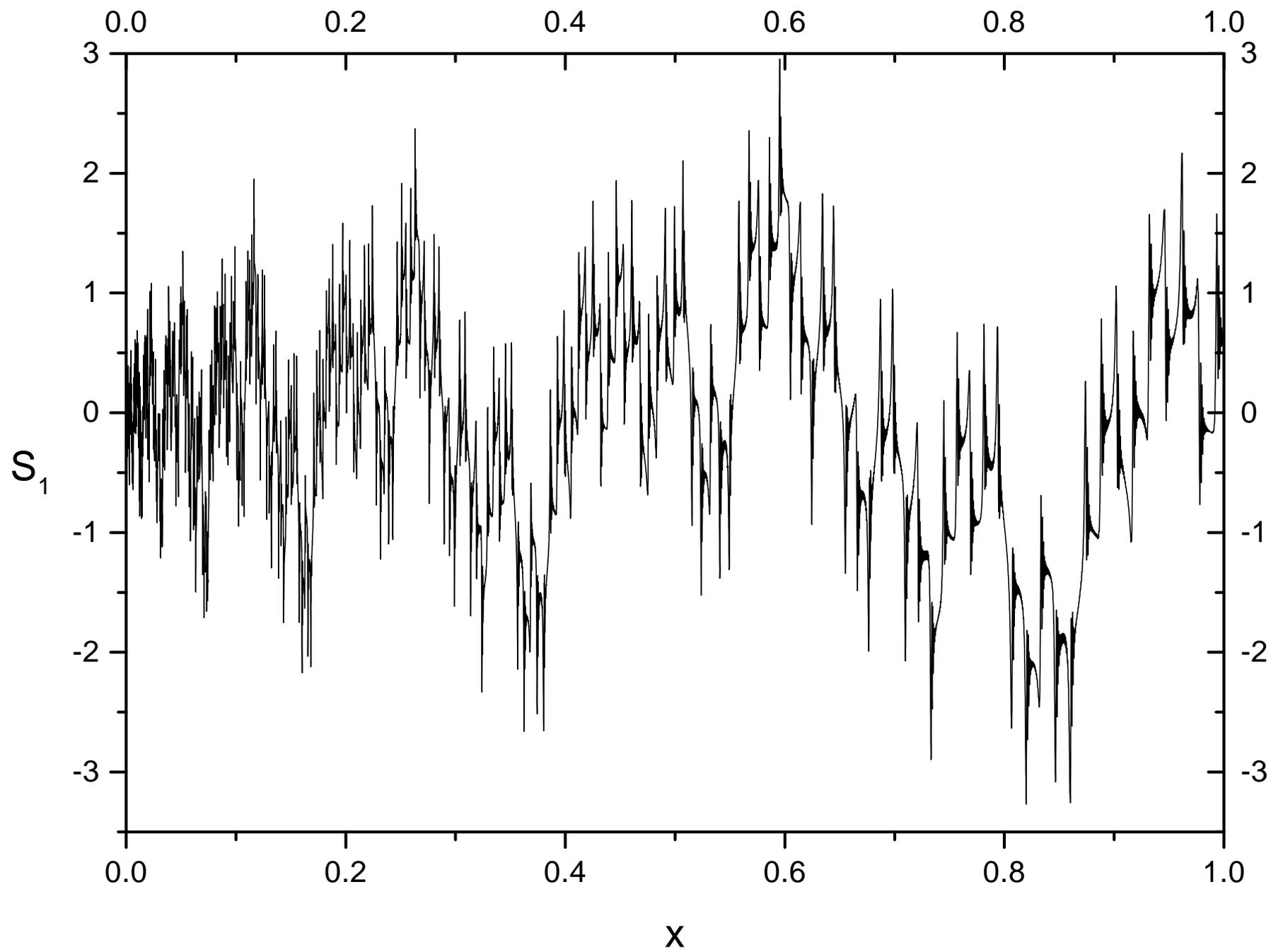
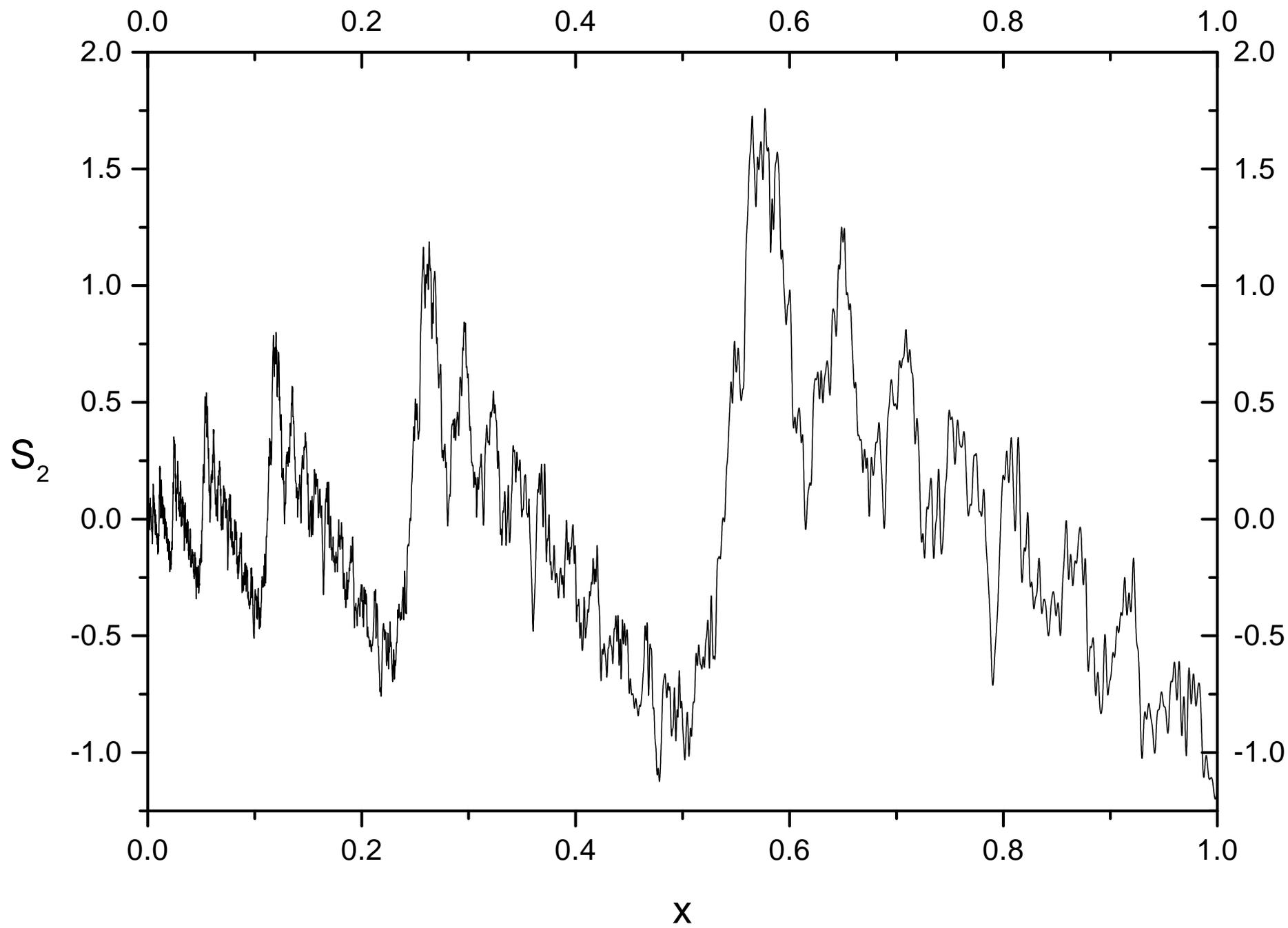
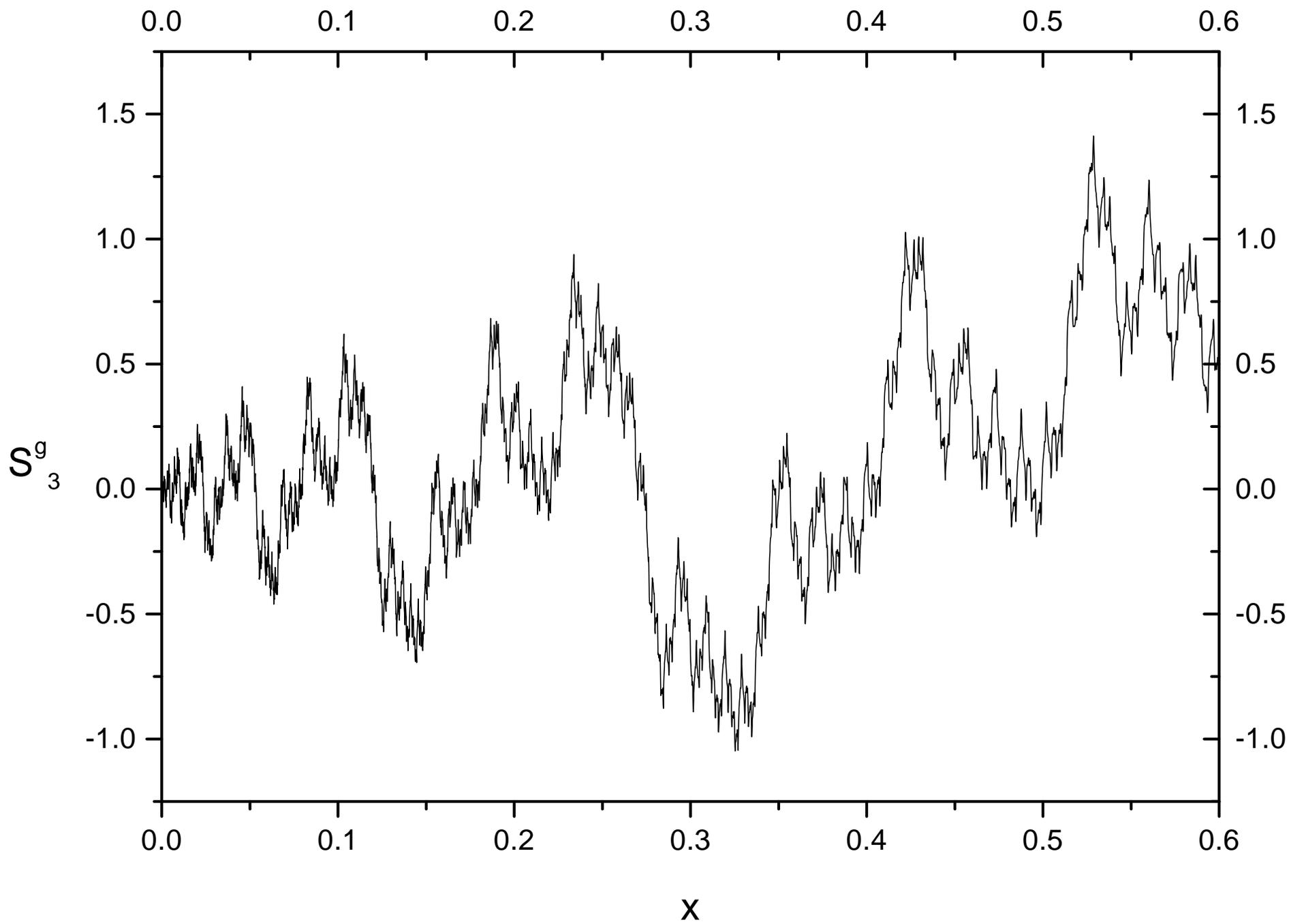
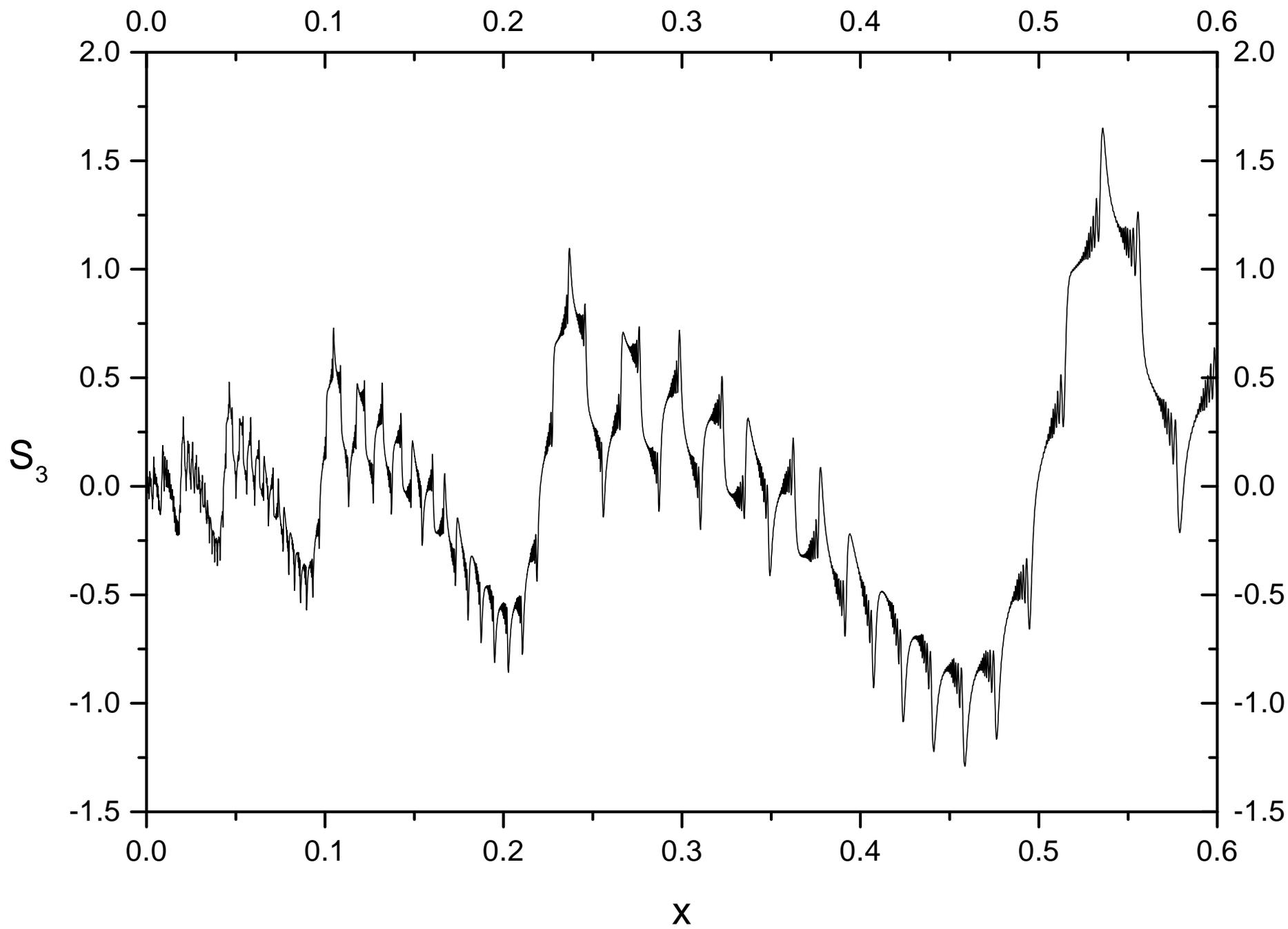
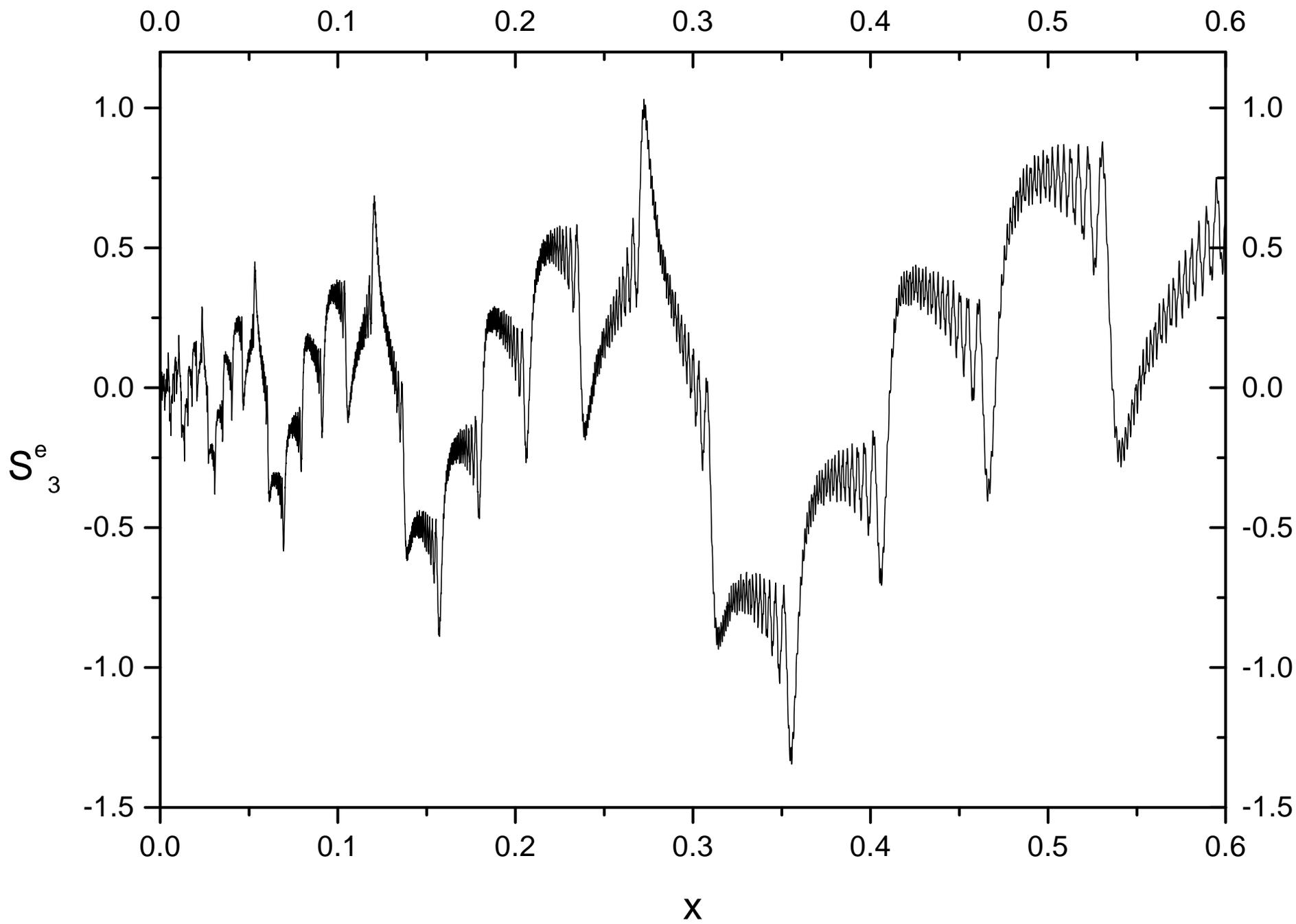
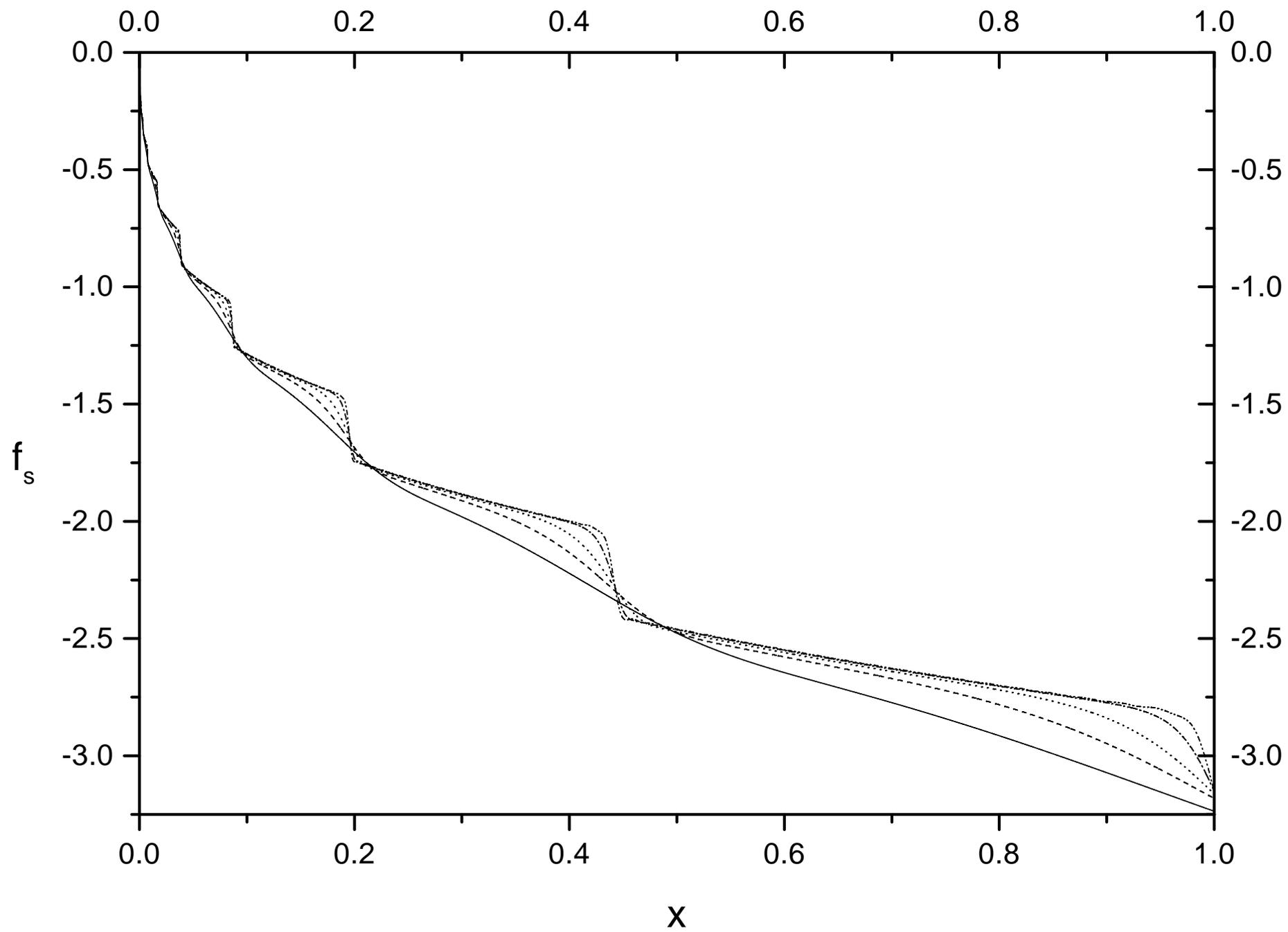
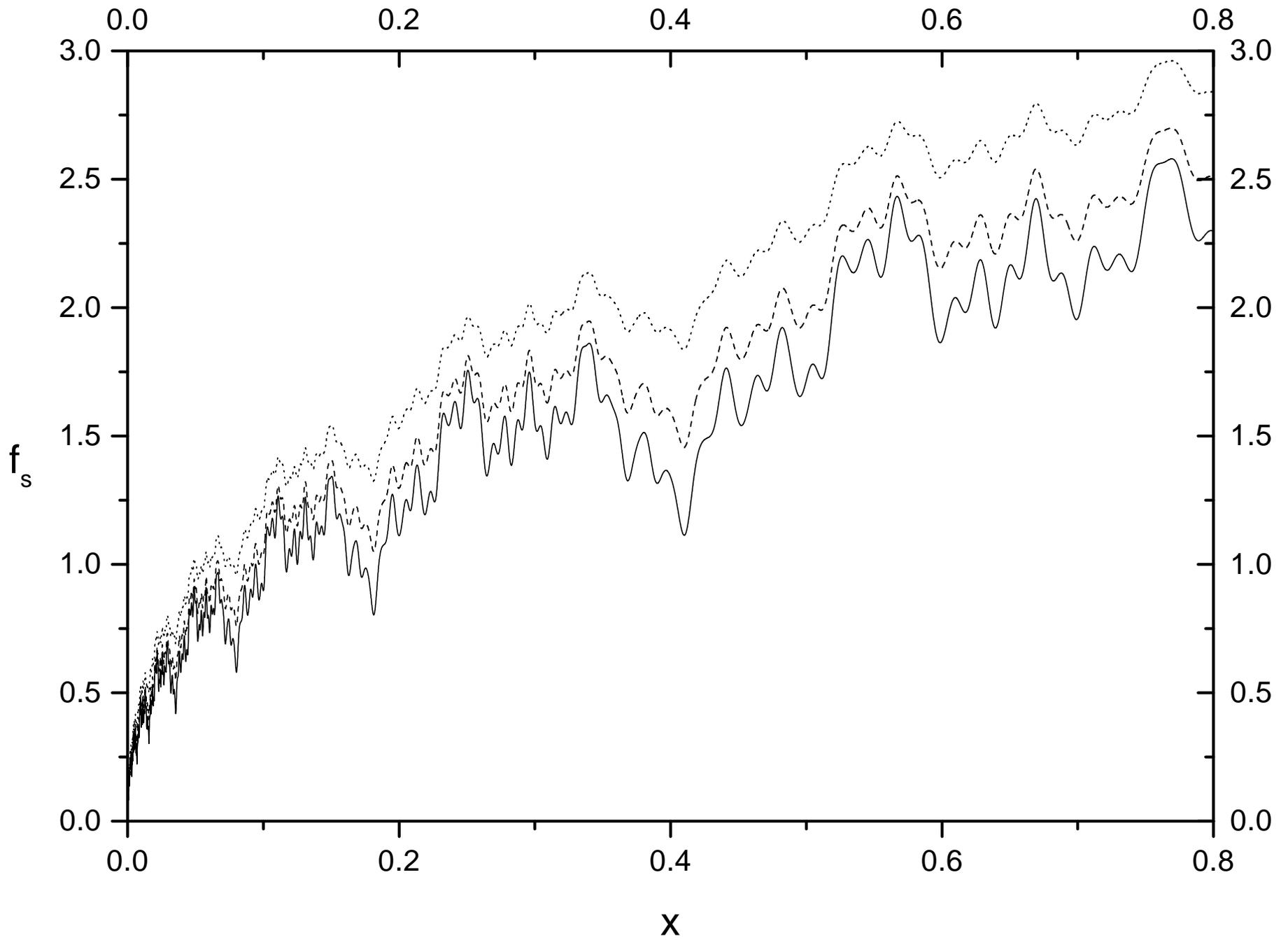
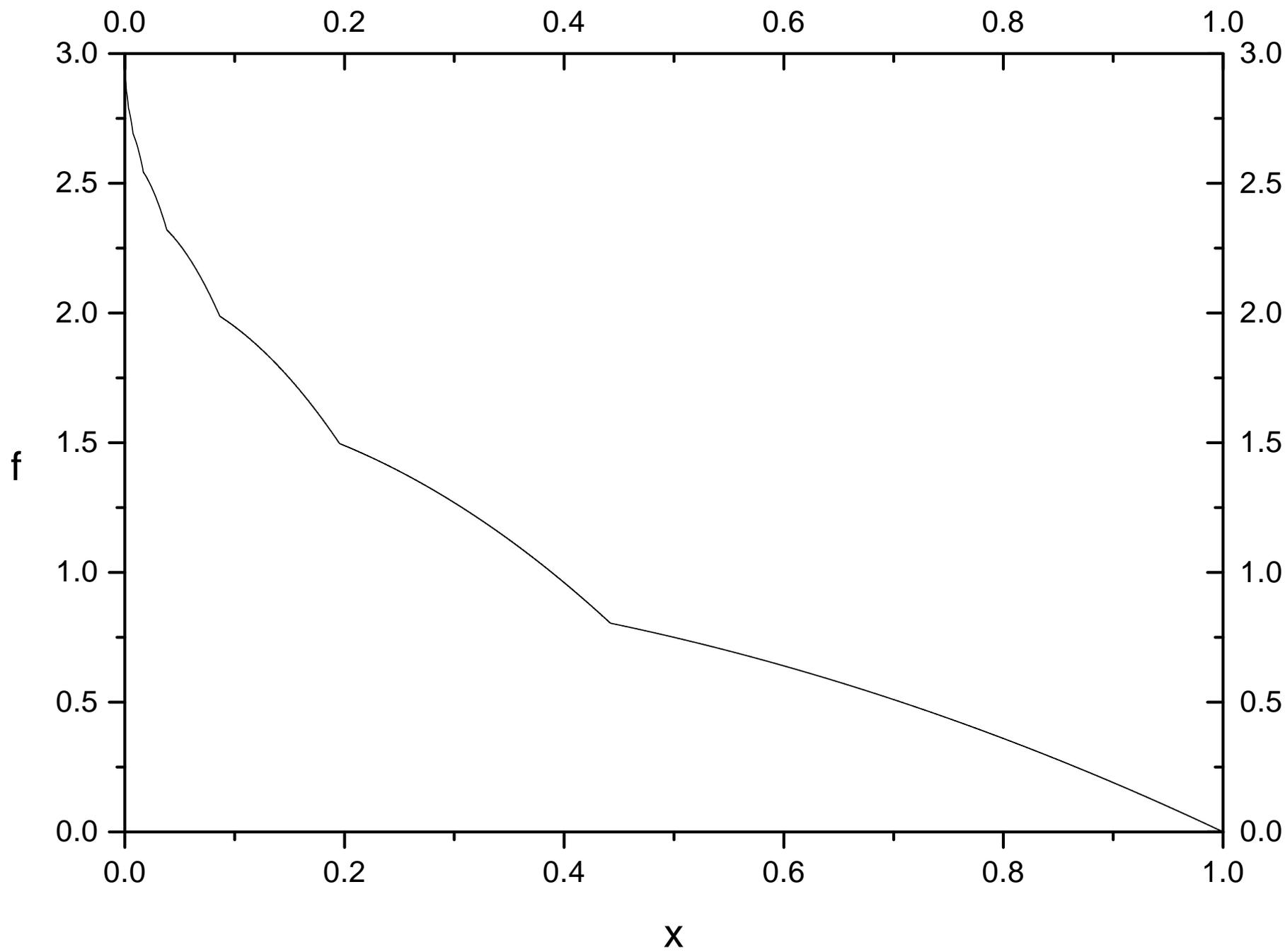
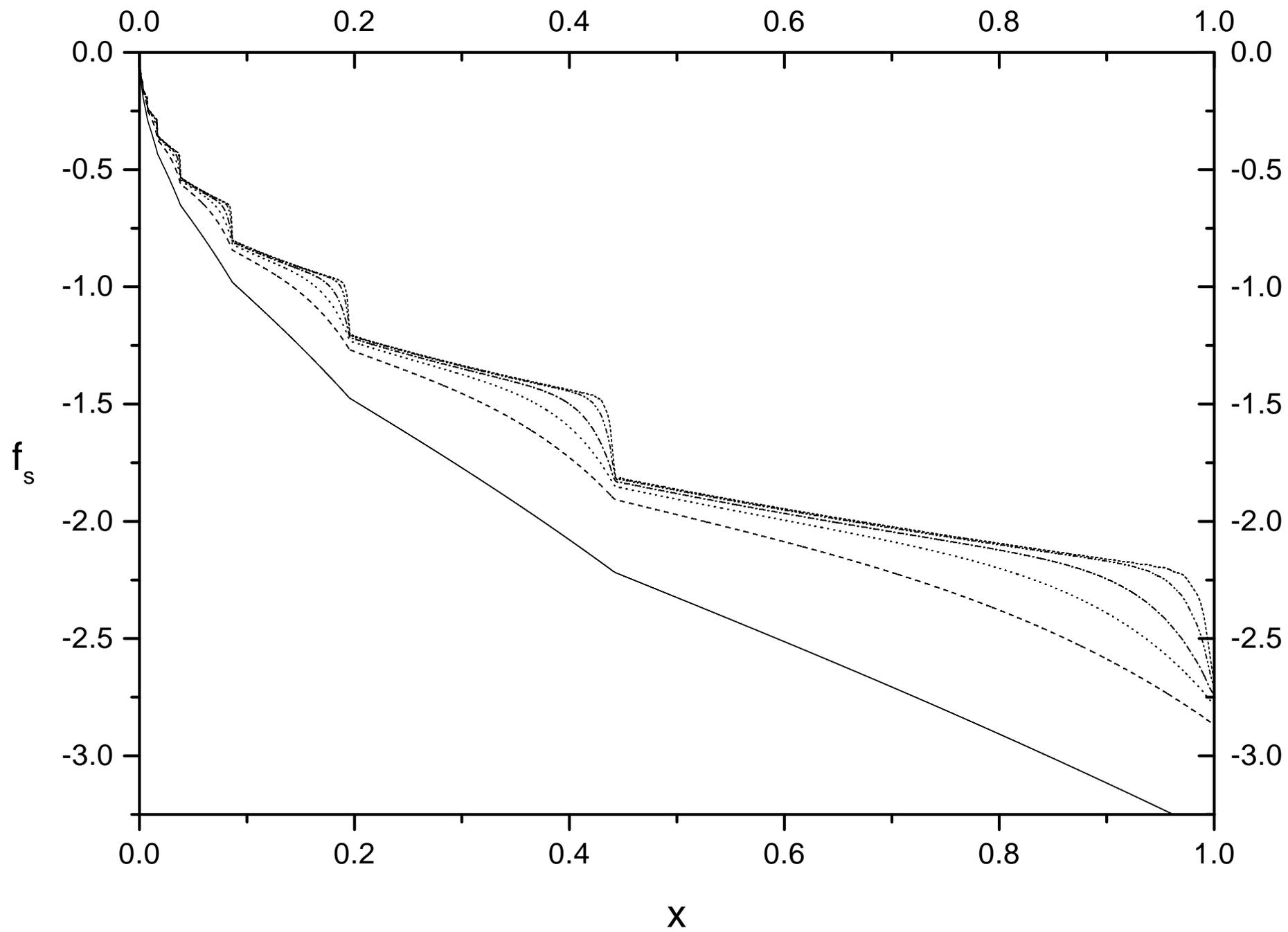
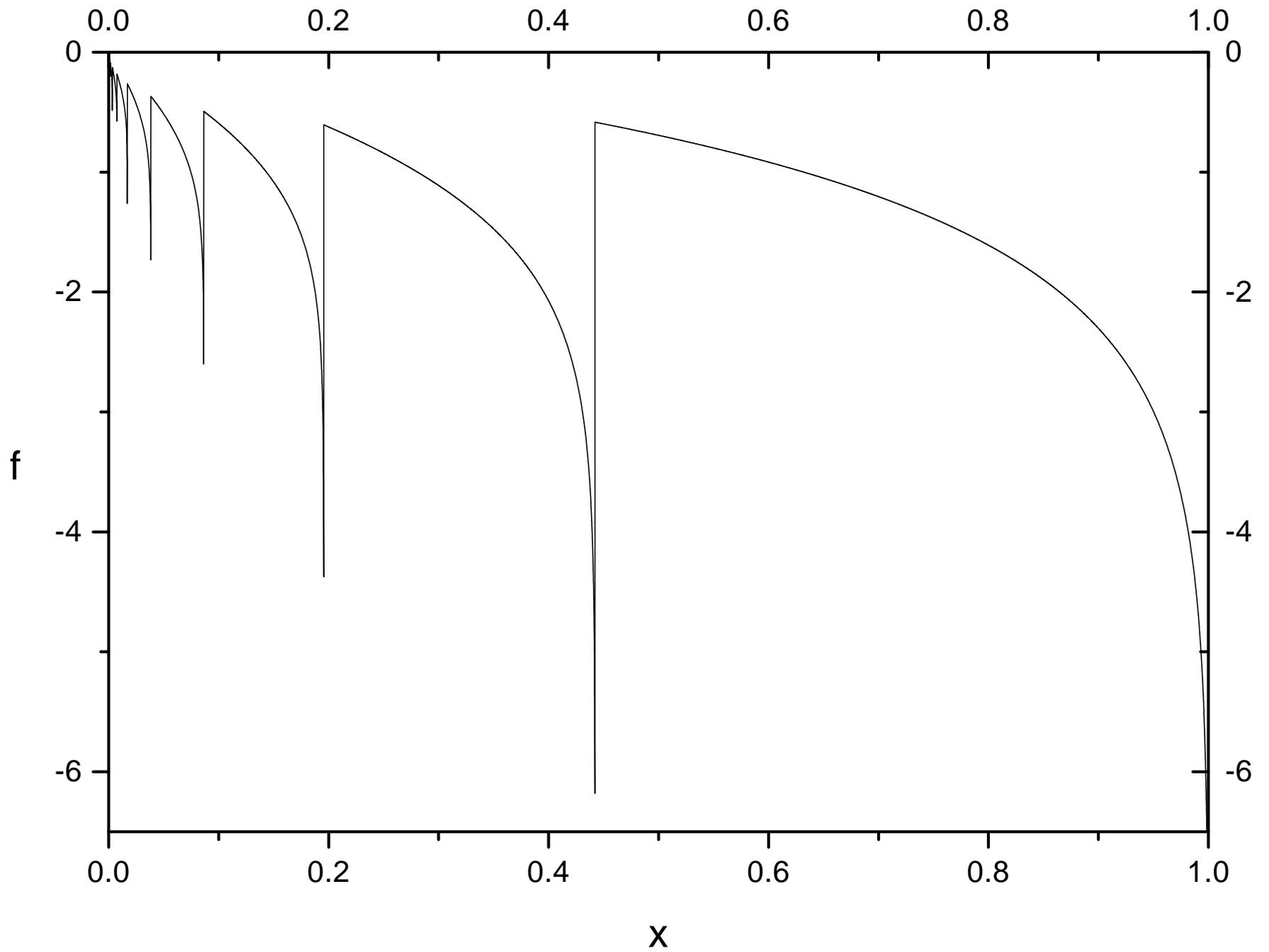
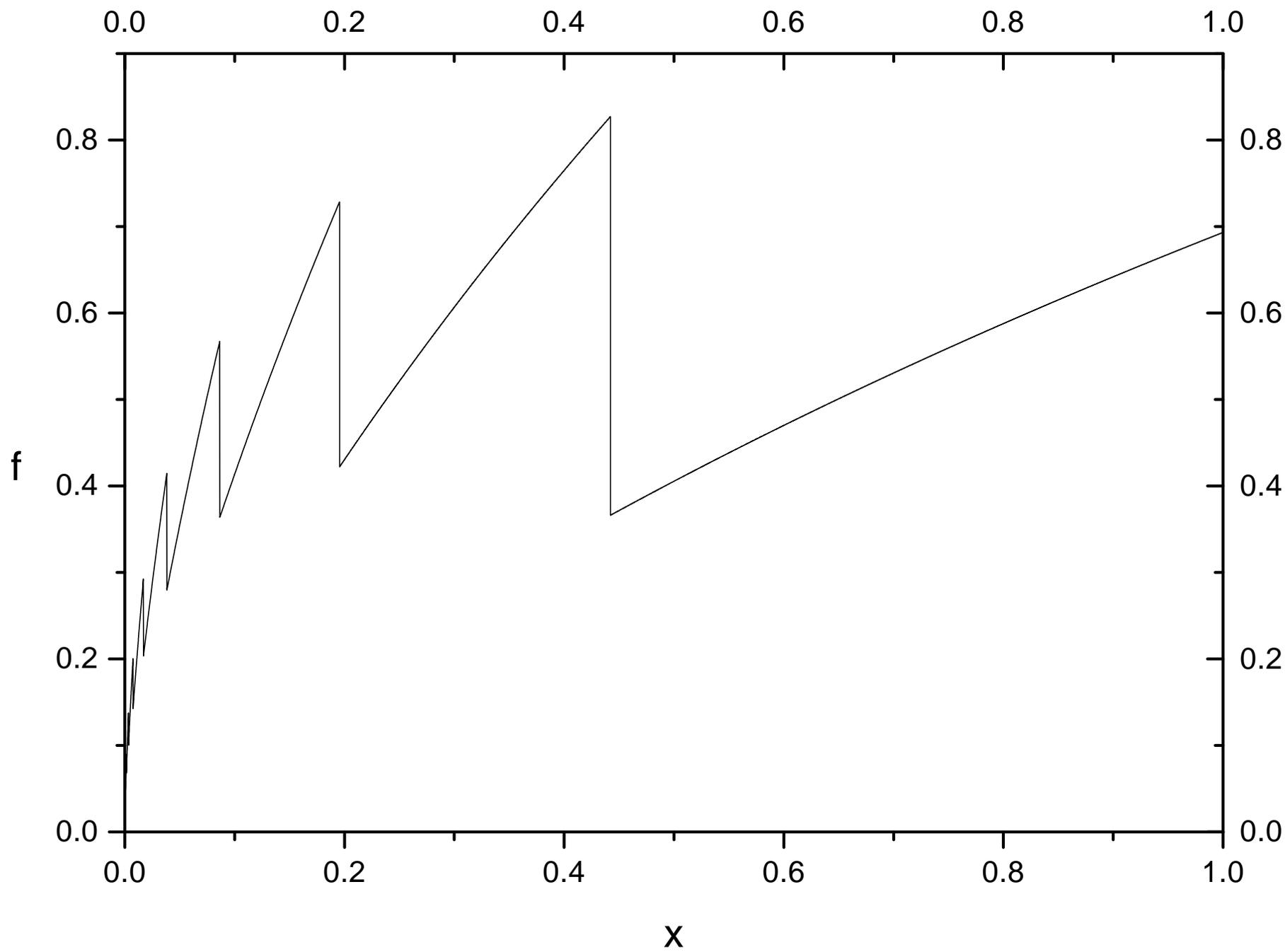
\end{document}